\documentclass[prd,twocolumn,floatfix,amsmath,nofootinbib,amssymb,floatfix]{revtex4}
\usepackage{graphicx,color,dcolumn,booktabs,bm}
\usepackage{longtable,lscape}
\usepackage{pdfpages}
\usepackage{txfonts}
\usepackage{overpic}
\usepackage{amssymb}
\usepackage{makecell}
\usepackage{indentfirst}
\usepackage{feynmf}   
\usepackage{slashed}  
\usepackage{cases}
\usepackage{color}
\usepackage{multirow}
\usepackage{threeparttable}
\usepackage{epstopdf}
\usepackage{enumerate}
\usepackage{diagbox}
\usepackage{graphicx,color,dcolumn,booktabs,bm}
\usepackage[colorlinks,
            citecolor=blue,
            anchorcolor=red,
            menucolor=red,
            linkcolor=red,
            filecolor=red,
            runcolor=red,
            urlcolor=blue,
            frenchlinks=red]{hyperref}
\usepackage{tikz}

\begin{document}

\title{Prediction of $QQqq\bar{s}$ molecular pentaquarks within the extended local hidden gauge approach}

\author{Zhong-Yu Wang$^{1}$}
\email{zhongyuwang@gzu.edu.cn}

\author{Zheng-Wen Long$^{1}$\footnote{Corresponding author}}
\email{zwlong@gzu.edu.cn}

\affiliation{
$^1$College of Physics, Guizhou University, Guiyang 550025, China
}

\date{\today}

\begin{abstract}

We investigate hadronic molecular states with the quark contents $ccqq\bar{s}$, $bbqq\bar{s}$, and $bcqq\bar{s}$ $(q=u,d)$ by employing the extended local hidden gauge approach.
Considering that the $S$-wave meson-baryon interactions are dominated by vector meson exchange, the coupled channels scattering amplitudes are obtained by solving the Bethe-Salpeter equation in its on-shell form.
We find that the poles appearing on the complex Riemann sheet are potential candidates for dynamically generated molecular pentaquark states.
The results suggest the existence of a total of fourteen molecular states with quantum numbers $I(J^{P})=0(1/2^{-})$, $0(3/2^{-})$, and $0(5/2^{-})$, which arise from the interactions of the $K^{(*)}\Xi_{cc}^{(*)}$, $K^{(*)}\Xi_{bb}^{(*)}$, $K^{(*)}\Xi_{bc}^{(*)}$, and $K^{(*)}\Xi_{bc}^{'}$ channels, respectively.
Their binding energies are calculated to be about $0.1-33$ MeV, and this range depends on the free parameter of the theory.
Our research contributes to the spectroscopic studies of hadronic molecular pentaquark states.

\end{abstract}
\maketitle

\section{Introduction}\label{sec:Introduction}

Current combined efforts at hadron facilities have led to the discovery of numerous new states containing heavy quarks, as summarized in reviews \cite{Liu:2013waa,Hosaka:2016pey,Chen:2016qju,Richard:2016eis,Lebed:2016hpi,Olsen:2017bmm,Guo:2017jvc,Liu:2019zoy,Brambilla:2019esw,Meng:2022ozq,Chen:2022asf,Liu:2024uxn}.
Some of these states exhibit inherently exotic properties that are cannot be explained within the conventional frameworks used to describe meson $q\bar{q}$ structures or baryon $qqq$ configurations.
Notably, a significant phenomenon is observed: the masses of many newly discovered hadron states are close to the thresholds of two specific hadron systems.
Consequently, these states are considered as strong candidates for hadronic molecular states.
For instance, the LHCb Collaboration discovered three hidden-charm pentaquark states, $P_{c}(4312)^{+}$, $P_{c}(4440)^{+}$, and $P_{c}(4457)^{+}$ in the decay process $\Lambda_{b}^{0} \rightarrow J/\psi pK^{-}$ \cite{LHCb:2019kea}.
There is evidence suggesting that they are hadronic molecular states arising from $\bar{D}^{(*)}\Sigma_{c}$ interactions \cite{Wu:2010jy,Wu:2010vk,Chen:2019asm,Chen:2019bip,Liu:2019tjn,He:2019ify,Xiao:2019mvs,Guo:2019kdc,Xiao:2019aya}.
In addition, the LHCb Collaboration observed the $P_{cs}(4459)^{0}$ state in the decay $\Xi_{b}^{-} \rightarrow J/\psi\Lambda K^{-}$ \cite{LHCb:2020jpq} and the $P_{cs}(4338)^{0}$ state in the decay $B^{-}\rightarrow J/\psi\Lambda\bar{p}$ \cite{LHCb:2022ogu}.
They are classified as molecular state candidates for the $\bar{D}^{*}\Xi_{c}$ and $\bar{D}\Xi_{c}$ systems because their masses are close to the corresponding thresholds \cite{Wu:2010jy,Wu:2010vk,Xiao:2021rgp,Du:2021bgb,Zhu:2021lhd,Lu:2021irg,Zou:2021sha,Wang:2021itn,Wu:2021caw,Yan:2022wuz,Meng:2022wgl,Burns:2022uha,Ortega:2022uyu}.

Over the past two decades, experimentally discovered candidates for hadronic molecular states have been primarily concentrated in the realm of charm physics.
In the field of charm mesons, numerous theoretical studies have attempted to interpret the BaBar and CLEO experimental findings \cite{BaBar:2003oey,CLEO:2003ggt} associated with the single-charm $D_{s0}(2317)$ and $D_{s1}(2460)$ states as manifestations of $DK$ and $D^{*}K$ molecular configurations \cite{Barnes:2003dj,Chen:2004dy,Guo:2006fu,Guo:2006rp,Navarra:2015iea,Faessler:2007us,Xie:2010zza,Feng:2012zze,Yang:2021tvc}.
In 2020, the LHCb Collaboration observed the $X_{0}(2900)$ and $X_{1}(2900)$ states \cite{LHCb:2020pxc}, which have been proposed as good candidates for the $\bar{D}^{*}K^{*}$ and $\bar{D}_{1}K$ molecular states, respectively, due to their proximity to the corresponding thresholds of these two channels \cite{Molina:2020hde,Kong:2021ohg,Wang:2021lwy,Chen:2020aos,Agaev:2020nrc,He:2020btl,Qi:2021iyv,Chen:2021tad}.
Subsequently, the LHCb Collaboration reported two single-charm tetraquark structures $T_{c\bar{s}}(2900)$ in the $D_{s}\pi$ invariant mass distributions in the process of $B$ meson decays \cite{LHCb:2022sfr,LHCb:2022lzp}.
It is reasonable to classify them as $D^{*}K^{*}$ molecular states based on their masses, quantum numbers, and peak structures \cite{Molina:2010tx,Agaev:2022duz,Agaev:2022eyk,Duan:2023qsg,Wang:2023hpp}.
Last year, the LHCb Collaboration observed two new similar $T_{c\bar{s}}$ states in the $D_{s}^{+}\pi^{+}$ and $D_{s}^{+}\pi^{-}$ invariant mass spectra of the $D_{s1}(2460)^{+} \rightarrow D_{s}^{+}\pi^{+}\pi^{-}$ decay \cite{LHCb:2024iuo}.
Reference \cite{Wang:2024fsz} demonstrated that the $S$-wave scattering amplitude of the $D_{s}\pi$ and $DK$ coupled channels system can dynamically generate these two new $T_{c\bar{s}}$ tetraquark states, and that the amplitude based on the final state interaction provides a good description of the experimental data from Ref. \cite{LHCb:2024iuo}.
For the double-charm tetraquark state, it is well known that the $T_{cc}(3875)^{+}$ state was discovered by the LHCb Collaboration in the $D^{0}D^{0}\pi^{+}$ invariant mass distribution \cite{LHCb:2021vvq}, and which is a good candidate for a $DD^{*}$ doubly charmed molecular state \cite{Manohar:1992nd,Ericson:1993wy,Tornqvist:1993ng,Janc:2004qn,Ding:2009vj,Ohkoda:2012hv,Li:2012ss,Xu:2017tsr,Liu:2019stu,Tang:2019nwv,Ding:2020dio,Feijoo:2021ppq}.

In the field of charm baryons, a series of candidates for singly charmed hadronic molecular states have been experimentally observed.
The BaBar Collaboration first observed the $\Lambda_{c}(2940)^{+}$ state in the $D^{0}p$ invariant mass distribution in Ref. \cite{BaBar:2006itc}, an observation that was confirmed in the $\pi\Sigma_{c}(2455)$ decay channel by the Belle Collaboration \cite{Belle:2006xni}.
The Belle Collaboration reported the discovery of the $\Lambda_{c}(2910)^{+}$ particle in the $\pi\Sigma_{c}(2455)$ invariant mass distribution in 2023 \cite{Belle:2022hnm}.
The $\Lambda_{c}(2940)^{+}$ and $\Lambda_{c}(2910)^{+}$ states are widely discussed as the molecular states of $D^{*}N$ in Refs. \cite{He:2006is,Garcia-Recio:2008rjt,He:2010zq,Dong:2009tg,Dong:2010xv,Liang:2011zza,Ortega:2012cx,Zhang:2012jk,Ortega:2013fta,Dong:2014ksa,Ortega:2014eoa,Xie:2015zga,Yang:2015eoa,Zhao:2016zhf,Zhang:2019vqe,Wang:2020dhf,Xin:2023gkf,Yan:2023ttx,Yue:2024paz}.
In addition, the LHCb and Belle Collaborations discovered several narrow $\Omega_{c}^{0}$ states that decay into the $\Xi_{c}^{+}K^{-}$ channel in Refs. \cite{LHCb:2017uwr,Belle:2017ext,LHCb:2021ptx,LHCb:2023sxp}.
Among them, the $\Omega_{c}(3050)^{0}$ and $\Omega_{c}(3090)^{0}$ states are closely related to the $\Xi_{c}\bar{K}$ meson-baryon coupled channels interaction, as discussed in Refs. \cite{Debastiani:2017ewu,Montana:2017kjw,Debastiani:2018adr}.
However, to date, no experimental candidates corresponding to the double-charm pentaquark state have been conclusively identified through experimental investigations.
Fortunately, the LHCb Collaboration is actively investigating additional doubly heavy baryons, such as $\Omega_{cc}$, $\Xi_{bc}$, and $\Omega_{bc}$ \cite{LHCb:2021xba}.

The study of double-charm pentaquark molecular states has been a central focus of theoretical research.
In Refs. \cite{Hofmann:2005sw,Romanets:2012hm,Dias:2018qhp,Yan:2018zdt,Dong:2021bvy,Wang:2022aga,Wang:2023mdj}, the spectroscopic properties of $\Xi_{cc}$- and $\Omega_{cc}$-type pentaquark states were investigated through the $S$-wave meson-baryon interaction framework by solving the Bethe-Salpeter equation within the on-shell approximation.
In Refs. \cite{Chen:2021kad,Wang:2023aob,Yalikun:2023waw,Wang:2023ael,Sheng:2024hkf}, the mass spectra of doubly charmed molecular pentaquark candidates, including the $D^{(*)}\Sigma_{c}^{(*)}$, $D_{s}^{(*)}\Omega_{c}^{(*)}$, $D_{s}^{(*)}\Xi_{c}^{(',*)}$, $D^{(*)}\Xi_{c}^{(',*)}$, $K^{*}\Xi_{cc}/\bar{K}^{*}\Xi_{cc}$, and other systems were predicted using the one-boson-exchange model.
The purpose of our work is to enhance the understanding of the mass spectrum of double-heavy pentaquark molecular states and to provide more information for experimental exploration.

In this paper, we explore the possible molecular pentaquark state candidates with the quark contents $ccqq\bar{s}$, $bbqq\bar{s}$, and $bcqq\bar{s}$ within the extended local hidden gauge approach.
In this approach, the meson-baryon interactions are dominated by the exchange of a vector meson.
Using the $S$-wave interaction potential as the kernel, we calculate the scattering amplitude by solving the coupled channels Bethe-Salpeter equation on-shell form to find the dynamically generate molecular states.
We will determine their masses and widths by analyzing the pole positions on the complex Riemann sheet and evaluate their couplings to each channel to determine the dominant one.
For the $ccqq\bar{s}$ system with isospin $I=0$, we consider the interactions inside the $K^{(*)}\Xi_{cc}^{(*)}$ and $D_{s}^{(*)}\Lambda_{c}$ channels.
For the $bbqq\bar{s}$ system with isospin $I=0$, we consider the interactions inside the $K^{(*)}\Xi_{bb}^{(*)}$ and $\bar{B}_{s}^{(*)}\Lambda_{b}$ channels.
For the $bcqq\bar{s}$ system with isospin $I=0$, we consider the interactions inside the $K^{(*)}\Xi_{bc}^{(',*)}$, $D_{s}^{(*)}\Lambda_{b}$, and $\bar{B}_{s}^{(*)}\Lambda_{c}$ channels.

The outline of the paper is as follows. 
In Sec. \ref{sec:Formalism}, we first present the transition potentials arising from the meson-baryon interactions within the extended local hidden gauge approach, and then proceed to introduce the scattering amplitudes obtained from the Bethe-Salpeter equation.
In Sec. \ref{sec:Results}, we extract both the poles and their couplings to each channel from the coupled channels amplitudes.
A brief summary is given in Sec. \ref{sec:Summary}.

\section{Formalism}\label{sec:Formalism}

In this section, we first construct the coupled channels systems that contain the quark contents $ccqq\bar{s}$ ($q=u,d$), $bbqq\bar{s}$, and $bcqq\bar{s}$, as listed in Table \ref{tab:coupledchannels}.
Only systems with isospin $I = 0$ are considered in this study, because interactions between those with $I = 1$ are repulsive, and therefore, no bound states are expected to form.
Following Ref. \cite{Wang:2024yjp}, we separate the coupled channels into four blocks: $PB(1/2^{+})$, $PB(3/2^{+})$, $VB(1/2^{+})$, and $VB(3/2^{+})$, where the symbol $P$ represents a pseudoscalar meson, $V$ represents a vector meson, $B(1/2^{+})$ denotes the ground baryon with spin and parity $J^{P}=1/2^{+}$, and $B(3/2^{+})$ denotes the ground baryon with $J^{P}=3/2^{+}$.
In addition, the thresholds for these coupled channels are listed in Table \ref{tab:Thresholds}, where some particle masses are taken from the Particle Data Group (PDG) \cite{ParticleDataGroup:2024cfk}, while others are the theoretical predictions provided by the constituent quark model in Refs. \cite{Roncaglia:1995az,Roberts:2007ni,Lu:2017meb,Weng:2018mmf}.

\begin{table}[htbp]
\centering
\renewcommand\tabcolsep{4.5mm}
\renewcommand{\arraystretch}{1.50}
\caption{The coupled channels in different sectors are considered. The symbols $P$ and $V$ stand for the pseudoscalar and vector mesons, respectively, while $B(1/2^{+})$ and $B(3/2^{+})$ stand for the ground state baryons with spin and parity $J^{P}=1/2^{+}$ and $J^{P}=3/2^{+}$, respectively.}
\begin{tabular*}{86mm}{l|cccc}
\toprule[1.00pt]
\toprule[1.00pt]
&\multicolumn{2}{c|}{\mbox{$ccqq\bar{s}$}}&\multicolumn{2}{c}{\mbox{$bbqq\bar{s}$}}\\
\hline
$PB(1/2^{+})$&$K\Xi_{cc}$&\multicolumn{1}{c|}{$D_{s}\Lambda_{c}$}&$K\Xi_{bb}$&$\bar{B}_{s}\Lambda_{b}$\\
$PB(3/2^{+})$&$K\Xi_{cc}^{*}$&\multicolumn{1}{c|}{}&$K\Xi_{bb}^{*}$&\\
$VB(1/2^{+})$&$K^{*}\Xi_{cc}$&\multicolumn{1}{c|}{$D_{s}^{*}\Lambda_{c}$}&$K^{*}\Xi_{bb}$&$\bar{B}_{s}^{*}\Lambda_{b}$\\
$VB(3/2^{+})$&$K^{*}\Xi_{cc}^{*}$&\multicolumn{1}{c|}{}&$K^{*}\Xi_{bb}^{*}$&\\
\hline
\multicolumn{1}{c}{}&\multicolumn{4}{c}{\mbox{$bcqq\bar{s}$}}\\
\hline
$PB(1/2^{+})$&$K\Xi_{bc}$&$K\Xi_{bc}^{'}$&$D_{s}\Lambda_{b}$&$\bar{B}_{s}\Lambda_{c}$\\
$PB(3/2^{+})$&$K\Xi_{bc}^{*}$&&&\\
$VB(1/2^{+})$&$K^{*}\Xi_{bc}$&$K^{*}\Xi_{bc}^{'}$&$D_{s}^{*}\Lambda_{b}$&$\bar{B}_{s}^{*}\Lambda_{c}$\\
$VB(3/2^{+})$&$K^{*}\Xi_{bc}^{*}$&&&\\
\bottomrule[1.00pt]
\bottomrule[1.00pt]
\end{tabular*}
\label{tab:coupledchannels}
\end{table}

\begin{table*}[htbp]
\centering
\renewcommand\tabcolsep{2.0mm}
\renewcommand{\arraystretch}{1.50}
\caption{Thresholds (in MeV) for the relevant reaction channels.}
\begin{tabular*}{178mm}{@{\extracolsep{\fill}}ccccccc}
\toprule[1.00pt]
\toprule[1.00pt]
\multicolumn{7}{c}{$ccqq\bar{s}$} \\
\hline
Channels&$K\Xi_{cc}$&$D_{s}\Lambda_{c}$&$K\Xi_{cc}^{*}$&$K^{*}\Xi_{cc}$&$D_{s}^{*}\Lambda_{c}$&$K^{*}\Xi_{cc}^{*}$ \\
Thresholds&$4117.64$&$4254.81$&$4170.64$&$4515.61$&$4398.66$&$4568.61$ \\
\hline
\multicolumn{7}{c}{$bbqq\bar{s}$} \\
\hline
Channels&$K\Xi_{bb}$&$\bar{B}_{s}\Lambda_{b}$&$K\Xi_{bb}^{*}$&$K^{*}\Xi_{bb}$&$\bar{B}_{s}^{*}\Lambda_{b}$&$K^{*}\Xi_{bb}^{*}$ \\
Thresholds&$10835.64$&$10986.53$&$10865.64$&$11233.61$&$11035.00$&$11263.61$ \\
\hline
\multicolumn{7}{c}{$bcqq\bar{s}$} \\
\hline
Channels&$K\Xi_{bc}$&$K\Xi_{bc}^{'}$&$D_{s}\Lambda_{b}$&$\bar{B}_{s}\Lambda_{c}$&$K\Xi_{bc}^{*}$& \\
Thresholds&$7417.64$&$7443.64$&$7587.95$&$7653.39$&$7468.64$& \\
Channels&$K^{*}\Xi_{bc}$&$K^{*}\Xi_{bc}^{'}$&$D_{s}^{*}\Lambda_{b}$&$\bar{B}_{s}^{*}\Lambda_{c}$&$K^{*}\Xi_{bc}^{*}$& \\
Thresholds&$7815.61$&$7841.61$&$7731.80$&$7701.86$&$7866.61$& \\
\bottomrule[1.00pt]
\bottomrule[1.00pt]
\end{tabular*}
\label{tab:Thresholds}
\end{table*}

\begin{figure}[htbp]
\centering
\includegraphics[width=0.8\linewidth,trim=150 580 250 120,clip]{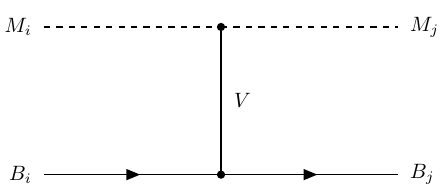}
\caption{The diagram illustrates the vector meson exchange mechanism in the meson-baryon interaction, where $M_{i}(M_{j})$ and $B_{i}(B_{j})$ represent the initial (final) meson and baryon, respectively, with $V$ denoting the exchanged vector meson.}
\label{fig:interaction}
\end{figure}

For the interactions in meson-baryon systems, we consider the mechanism that is dominated by vector meson exchange, as illustrated in Fig. \ref{fig:interaction}.
Therefore, the three typical interaction vertices of vector-pseudoscalar-pseudoscalar ($VPP$), three-vector-meson ($VVV$), and vector-baryon-baryon ($VBB$) are required.
The effective Lagrangians for these vertices can be derived using the extended local hidden gauge approach \cite{Liang:2017ejq,Yu:2018yxl,Yu:2019yfr,Dias:2019klk,Liang:2020dxr}, and they are given by

\begin{equation}
\begin{aligned}
\mathcal{L}_{V P P}=-i g\left\langle\left[P, \partial_\mu P\right] V^\mu\right\rangle,
\end{aligned}
\label{eq:LVVP}
\end{equation}
\begin{equation}
\begin{aligned}
\mathcal{L}_{V V V}=i g\left\langle\left(V^\mu \partial_\nu V_\mu-\partial_\nu V^\mu V_\mu\right) V^\nu\right\rangle,
\end{aligned}
\label{eq:LVVV}
\end{equation}
\begin{equation}
\begin{aligned}
\mathcal{L}_{V B B}=g\left(\left\langle\bar{B} \gamma_\mu\left[V^\mu, B\right]\right\rangle+\left\langle\bar{B} \gamma_\mu B\right\rangle\left\langle V^\mu\right\rangle\right),
\end{aligned}
\label{eq:LVBB}
\end{equation}
where $\left\langle \right\rangle$ stands for the matrix trace, $g$ is the coupling constant, which is given by $M_{V}/(2f_{\pi})$, where $M_{V}$ denotes the mass of the exchanged light vector meson and $f_{\pi}=93$ MeV is the pion decay constant.
The symbols $P$ and $V^{\mu}$ represent the pseudoscalar and vector meson matrices of SU(5), respectively, as shown below

\begin{equation}
\begin{aligned}
P=\left(\begin{array}{ccccc}
\frac{\eta}{\sqrt{3}}+\frac{\eta^{\prime}}{\sqrt{6}}+\frac{\pi^0}{\sqrt{2}} & \pi^{+} & K^{+} & \bar{D}^0 & B^{+} \\
\pi^{-} & \frac{\eta}{\sqrt{3}}+\frac{\eta^{\prime}}{\sqrt{6}}-\frac{\pi^0}{\sqrt{2}} & K^0 & D^{-} & B^0 \\
K^{-} & \bar{K}^0 & -\frac{\eta}{\sqrt{3}}+\sqrt{\frac{2}{3}}\eta^{\prime} & D_s^{-} & B_s^0 \\
D^0 & D^{+} & D_s^{+} & \eta_c & B_c^{+} \\
B^{-} & \bar{B}^0 & \bar{B}_s^0 & B_c^{-} & \eta_b
\end{array}\right),
\end{aligned}
\label{eq:P}
\end{equation}
\begin{equation}
\begin{aligned}
V^{\mu}=\left(\begin{array}{ccccc}
\frac{\omega+\rho^0}{\sqrt{2}} & \rho^{+} & K^{*+} & \bar{D}^{* 0} & B^{*+} \\
\rho^{-} & \frac{\omega-\rho^0}{\sqrt{2}} & K^{* 0} & D^{*-} & B^{* 0} \\
K^{*-} & \bar{K}^{* 0} & \phi & D_s^{*-} & B_s^{* 0} \\
D^{* 0} & D^{*+} & D_s^{*+} & J / \psi & B_c^{*+} \\
B^{*-} & \bar{B}^{* 0} & \bar{B}_s^{* 0} & B_c^{*-} & \Upsilon
\end{array}\right)^{\mu}.
\end{aligned}
\label{eq:V}
\end{equation}
Note that we exclusively considered the quark constituents of mesons and did not incorporate the SU(5) symmetry.
With Eqs. \eqref{eq:LVVP}, \eqref{eq:LVVV}, \eqref{eq:P}, and \eqref{eq:V}, we can easily obtain the interaction vertices $VPP$ and $VVV$ in Fig. \ref{fig:interaction}.
In Eq. \eqref{eq:LVBB}, the matrix $B$ corresponds to the baryon matrix within the SU(3) framework. 
However, extending this matrix to include the charm and bottom sectors presents significant challenges.
Therefore, following the approaches outlined in Refs. \cite{Debastiani:2017ewu,Dias:2018qhp}, we determine the $VBB$ vertices by employing both meson and baryon wave functions.
A more detailed discussion is provided in the Appendix of Ref. \cite{Debastiani:2017ewu}.
The spin-flavor wave functions of all baryons studied in this work are presented in Table \ref{tab:Wavefunctions}.
The flavor component of the wave function is explicitly constructed by leaving the heavy quarks as spectators and applying SU(3) symmetry to the light quarks, which may exhibit either mixed symmetric or mixed antisymmetric configurations.
The spin part can be either mixed symmetric $\chi_{MS}$, mixed antisymmetric $\chi_{MA}$, or fully symmetric $\chi_{S}$.
For the specific case of $S_{z}=+1/2$ in the context of the $J^{P}=1/2^{+}$ ground baryon states, the spin wave functions are given by
\begin{equation}
\begin{aligned}
\chi_{MS}(12)=\frac{1}{\sqrt{6}}(\uparrow\downarrow\uparrow+\downarrow\uparrow\uparrow-2\uparrow\uparrow\downarrow),
\end{aligned}
\label{eq:chiMS12}
\end{equation}
\begin{equation}
\begin{aligned}
\chi_{MS}(23)=\frac{1}{\sqrt{6}}(\uparrow\downarrow\uparrow+\uparrow\uparrow\downarrow-2\downarrow\uparrow\uparrow),
\end{aligned}
\label{eq:chiMS23}
\end{equation}
\begin{equation}
\begin{aligned}
\chi_{MA}(23)=\frac{1}{\sqrt{2}}(\uparrow\uparrow\downarrow-\uparrow\downarrow\uparrow).
\end{aligned}
\label{eq:chiMA23}
\end{equation}

\begin{table}[htbp]
\centering
\renewcommand\tabcolsep{2mm}
\renewcommand{\arraystretch}{1.50}
\caption{The wave functions for baryons with isospin and spin-parity quantum numbers $I(J^{P})=0(1/2^{+})$, $1/2(1/2^{+})$, and $1/2(3/2^{+})$, where $\chi_{MS}$, $\chi_{MA}$, and $\chi_{S}$ denote mixed symmetric, mixed antisymmetric, and fully symmetric spin wave functions, respectively.}
\begin{tabular*}{86mm}{@{\extracolsep{\fill}}lccc}
\toprule[1.00pt]
\toprule[1.00pt]
States&$I(J^{P})$&Flavor& Spin\\
\hline
$\Lambda_{c}^{+}$&$0(\frac{1}{2}^{+})$&$\frac{1}{\sqrt{2}}c(ud-du)$&$\chi_{MA}(23)$\\
$\Xi_{cc}^{++}$&$\frac{1}{2}(\frac{1}{2}^{+})$&$ccu$&$\chi_{MS}(12)$\\
$\Xi_{cc}^{+}$&$\frac{1}{2}(\frac{1}{2}^{+})$&$ccd$&$\chi_{MS}(12)$\\
$\Lambda_{b}^{0}$&$0(\frac{1}{2}^{+})$&$\frac{1}{\sqrt{2}}b(ud-du)$&$\chi_{MA}(23)$\\
$\Xi_{bb}^{0}$&$\frac{1}{2}(\frac{1}{2}^{+})$&$bbu$&$\chi_{MS}(12)$\\
$\Xi_{bb}^{-}$&$\frac{1}{2}(\frac{1}{2}^{+})$&$bbd$&$\chi_{MS}(12)$\\
$\Xi_{bc}^{+}$&$\frac{1}{2}(\frac{1}{2}^{+})$&$\frac{1}{\sqrt{2}}b(cu-uc)$&$\chi_{MA}(23)$\\
$\Xi_{bc}^{0}$&$\frac{1}{2}(\frac{1}{2}^{+})$&$\frac{1}{\sqrt{2}}b(cd-dc)$&$\chi_{MA}(23)$\\
$\Xi_{bc}^{'+}$&$\frac{1}{2}(\frac{1}{2}^{+})$&$\frac{1}{\sqrt{2}}b(cu+uc)$&$\chi_{MS}(23)$\\
$\Xi_{bc}^{'0}$&$\frac{1}{2}(\frac{1}{2}^{+})$&$\frac{1}{\sqrt{2}}b(cd+dc)$&$\chi_{MS}(23)$\\
$\Xi_{cc}^{*++}$&$\frac{1}{2}(\frac{3}{2}^{+})$&$ccu$&$\chi_{S}$\\
$\Xi_{cc}^{*+}$&$\frac{1}{2}(\frac{3}{2}^{+})$&$ccd$&$\chi_{S}$\\
$\Xi_{bb}^{*0}$&$\frac{1}{2}(\frac{3}{2}^{+})$&$bbu$&$\chi_{S}$\\
$\Xi_{bb}^{*-}$&$\frac{1}{2}(\frac{3}{2}^{+})$&$bbd$&$\chi_{S}$\\
$\Xi_{bc}^{*+}$&$\frac{1}{2}(\frac{3}{2}^{+})$&$\frac{1}{\sqrt{2}}b(cu+uc)$&$\chi_{S}$\\
$\Xi_{bc}^{*0}$&$\frac{1}{2}(\frac{3}{2}^{+})$&$\frac{1}{\sqrt{2}}b(cd+dc)$&$\chi_{S}$\\
\bottomrule[1.00pt]
\bottomrule[1.00pt]
\end{tabular*}
\label{tab:Wavefunctions}
\end{table}

\noindent
In the case of $S_{z}=+3/2$ for the $J^{P}=3/2^{+}$ ground baryon states, we have
\begin{equation}
\begin{aligned}
\chi_{S}=\uparrow\uparrow\uparrow.
\end{aligned}
\label{eq:chiS}
\end{equation}
The overlap of the spin wave functions that needs to be taken into account is as follows
\begin{equation}
\begin{aligned}
\left\langle\chi_{MS}(12)|\chi_{MS}(23)\right\rangle=-\frac{1}{2},
\end{aligned}
\label{eq:chiSchiS}
\end{equation}
\begin{equation}
\begin{aligned}
\left\langle\chi_{MS}(12)|\chi_{MA}(23)\right\rangle=-\frac{\sqrt{3}}{2}.
\end{aligned}
\label{eq:chiSchiA}
\end{equation}

For instance, consider a scenario where the $VBB$ vertex consists of a $\rho^{0}$ meson and two $\Xi_{cc}^{+}$ baryons, which corresponds to the $\Xi_{cc}^{+}\Xi_{cc}^{+}\rho^{0}$ vertex.
It appears during the process of the $K^{+}\Xi_{cc}^{+} \rightarrow K^{+}\Xi_{cc}^{+}$ transition.
The $\rho^{0}$ meson wave function, which is expressed in terms of quark constituents, can be defined as
\begin{equation}
\begin{aligned}
\rho^{0}=\frac{1}{\sqrt{2}}(u\bar{u}-d\bar{d}).
\end{aligned}
\label{eq:rho}
\end{equation}
With the spin-flavor wave functions provided in Table \ref{tab:Wavefunctions}, we can derive the $\Xi_{cc}^{+}\Xi_{cc}^{+}\rho^{0}$ vertex by evaluating the matrix element $\left\langle \Xi_{cc}^{+} \left| g\rho^{0} \right| \Xi_{cc}^{+} \right\rangle$, which is
\begin{equation}
\begin{aligned}
&\left\langle \Xi_{cc}^{+} \left| g\rho^{0} \right| \Xi_{cc}^{+} \right\rangle \\
&=g\left\langle ccd\otimes\chi_{MS}(12)\left|\left[\frac{1}{\sqrt{2}}(u\bar{u}-d\bar{d})\right]\right|\chi_{MS}(12)\otimes ccd\right\rangle \\
&=-\frac{g}{\sqrt{2}}.
\end{aligned}
\label{eq:XiccXiccrho}
\end{equation}
The approach employed in Refs. \cite{Debastiani:2017ewu,Montana:2017kjw} inherently neglects the three-momentum of the particles relative to the vector meson mass. 
This neglect enables the approximation $\gamma_{\mu} \rightarrow \gamma_{0}$, which in turn allows Eq. \eqref{eq:rho} to be applied within the $\Xi_{cc}^{+}$ baryon wave function as a number operator associated with a coupling strength $g$. 
This procedure results in a spin-independent operator at the quark level.
Finally, we can calculate all transition amplitudes by following the steps outlined in the Appendix of Ref. \cite{Debastiani:2017ewu}, as each transition has the same structural form, i.e.,
\begin{equation}
\begin{aligned}
v_{ij}=-C_{ij}\frac{1}{4f_{\pi}^{2}}(p_{i}^{0}+p_{j}^{0}),
\end{aligned}
\label{eq:vij1}
\end{equation}
where $p_{i}^{0}$ and $p_{j}^{0}$ represent the energies of the incoming meson and the outgoing meson, respectively, and $C_{ij}$ is the coefficient matrix listed in Tables \ref{tab:ccqqs1}-\ref{tab:bcqqs1}, which is symmetric.
The three parameters $\lambda_{c}$, $\lambda_{cc}$, and $\lambda_{b}$ in Tables \ref{tab:ccqqs1}-\ref{tab:bcqqs1} represent the suppression factors for the exchanged heavy vector mesons $\bar{D}^{*}$, $J/\psi$, and $B^{*}$, respectively, relative to the exchanged light vector mesons.
In our work, we adopt the parameter values $\lambda_{c}=1/4$, $\lambda_{cc}=1/9$, and $\lambda_{b}=1/10$ following previous studies in Refs. \cite{Debastiani:2017ewu,Dias:2019klk,Marse-Valera:2022khy}.
The contributions from heavier vector meson exchanges, such as those involving $B_{c}^{*}$ and $\Upsilon$ particles, have been omitted.
In addition, in the $VB \rightarrow VB$ process, we neglected the three-momenta of the exchanged vector mesons. 
Consequently, the polarization vectors of the external mesons satisfy $\epsilon_{1\mu} \epsilon_{3}^{\dagger\mu}=-\vec{\epsilon}_1 \cdot \vec{\epsilon}_3^{\dagger}$, where the temporal component is zero, that is, $\epsilon^{0}=0$.
Alternatively, we can use another expression that incorporates relativistic corrections to the $S$-wave in place of Eq. \eqref{eq:vij1}, as done in Ref. \cite{Oset:2001cn}
\begin{equation}
\begin{aligned}
v_{ij}(\sqrt{s})=-C_{ij}\frac{2\sqrt{s}-M_{i}-M_{j}}{4f_{\pi}^{2}}\left(\frac{M_{i}+E_{i}}{2M_{i}}\right)^{1/2}\left(\frac{M_{j}+E_{j}}{2M_{j}}\right)^{1/2},
\end{aligned}
\label{eq:vij2}
\end{equation}
where $M_{i}$ and $E_{i}$ are the mass and center-of-mass energy of the initial baryon, while $M_{j}$ and $E_{j}$ correspond to those of the final baryon, respectively.

\begin{table}[htbp]
\centering
\renewcommand\tabcolsep{0.5mm}
\renewcommand{\arraystretch}{1.50}
\caption{The coefficient matrix elements $C_{ij}$ in the $ccqq\bar{s}$ sector for the isospin $I=0$ configuration.}
\begin{tabular*}{86mm}{@{\extracolsep{\fill}}l|ccc|cc}
\toprule[1.00pt]
\toprule[1.00pt]
\multirow{3}{*}{$C_{ij}(PB(1/2^{+}),PB(3/2^{+}))$}&&$K\Xi_{cc}$&$D_{s}\Lambda_{c}$&&$K\Xi_{cc}^{*}$\\
&$K\Xi_{cc}$&$1$&$\frac{\sqrt{3}\lambda_{c}}{2}$&$K\Xi_{cc}^{*}$&$1$\\
&$D_{s}\Lambda_{c}$&$\frac{\sqrt{3}\lambda_{c}}{2}$&$-\lambda_{cc}$&&\\
\hline
\multirow{3}{*}{$C_{ij}(VB(1/2^{+}),VB(3/2^{+}))$}&&$K^{*}\Xi_{cc}$&$D_{s}^{*}\Lambda_{c}$&&$K^{*}\Xi_{cc}^{*}$\\
&$K^{*}\Xi_{cc}$&$1$&$\frac{\sqrt{3}\lambda_{c}}{2}$&$K^{*}\Xi_{cc}^{*}$&$1$\\
&$D_{s}^{*}\Lambda_{c}$&$\frac{\sqrt{3}\lambda_{c}}{2}$&$-\lambda_{cc}$&&\\
\bottomrule[1.00pt]
\bottomrule[1.00pt]
\end{tabular*}
\label{tab:ccqqs1}
\end{table}

\begin{table}[htbp]
\centering
\renewcommand\tabcolsep{0.5mm}
\renewcommand{\arraystretch}{1.50}
\caption{The coefficient matrix elements $C_{ij}$ in the $bbqq\bar{s}$ sector for the isospin $I=0$ configuration.}
\begin{tabular*}{86mm}{@{\extracolsep{\fill}}l|ccc|cc}
\toprule[1.00pt]
\toprule[1.00pt]
\multirow{3}{*}{$C_{ij}(PB(1/2^{+}),PB(3/2^{+}))$}&&$K\Xi_{bb}$&$\bar{B}_{s}\Lambda_{b}$&&$K\Xi_{bb}^{*}$\\
&$K\Xi_{bb}$&$1$&$\frac{\sqrt{3}\lambda_{b}}{2}$&$K\Xi_{bb}^{*}$&$1$\\
&$\bar{B}_{s}\Lambda_{b}$&$\frac{\sqrt{3}\lambda_{b}}{2}$&$0$&&\\
\hline
\multirow{3}{*}{$C_{ij}(VB(1/2^{+}),VB(3/2^{+}))$}&&$K^{*}\Xi_{bb}$&$\bar{B}_{s}^{*}\Lambda_{b}$&&$K^{*}\Xi_{bb}^{*}$\\
&$K^{*}\Xi_{bb}$&$1$&$\frac{\sqrt{3}\lambda_{b}}{2}$&$K^{*}\Xi_{bb}^{*}$&$1$\\
&$\bar{B}_{s}^{*}\Lambda_{b}$&$\frac{\sqrt{3}\lambda_{b}}{2}$&$0$&&\\
\bottomrule[1.00pt]
\bottomrule[1.00pt]
\end{tabular*}
\label{tab:bbqqs1}
\end{table}

\begin{table}[htbp]
\centering
\renewcommand\tabcolsep{2mm}
\renewcommand{\arraystretch}{1.50}
\caption{The coefficient matrix elements $C_{ij}$ in the $bcqq\bar{s}$ sector for the isospin $I=0$ configuration.}
\begin{tabular*}{86mm}{@{\extracolsep{\fill}}l|ccccc}
\toprule[1.00pt]
\toprule[1.00pt]
\multirow{5}{*}{$C_{ij}(PB(1/2^{+}))$}&&$K\Xi_{bc}$&$K\Xi_{bc}^{'}$&$D_{s}\Lambda_{b}$&$\bar{B}_{s}\Lambda_{c}$\\
&$K\Xi_{bc}$&$1$&$0$&$-\sqrt{2}\lambda_{c}$&$\frac{-\lambda_{b}}{2\sqrt{2}}$\\
&$K\Xi_{bc}^{'}$&$0$&$1$&$0$&$0$\\
&$D_{s}\Lambda_{b}$&$-\sqrt{2}\lambda_{c}$&$0$&$0$&$0$\\
&$\bar{B}_{s}\Lambda_{c}$&$\frac{-\lambda_{b}}{2\sqrt{2}}$&$0$&$0$&$0$\\
\hline
\multirow{2}{*}{$C_{ij}(PB(3/2^{+}))$}&&$K\Xi_{bc}^{*}$&&&\\
&$K\Xi_{bc}^{*}$&$1$&&&\\
\hline
\multirow{5}{*}{$C_{ij}(VB(1/2^{+}))$}&&$K^{*}\Xi_{bc}$&$K^{*}\Xi_{bc}^{'}$&$D_{s}^{*}\Lambda_{b}$&$\bar{B}_{s}^{*}\Lambda_{c}$\\
&$K^{*}\Xi_{bc}$&$1$&$0$&$-\sqrt{2}\lambda_{c}$&$\frac{-\lambda_{b}}{2\sqrt{2}}$\\
&$K^{*}\Xi_{bc}^{'}$&$0$&$1$&$0$&$0$\\
&$D_{s}^{*}\Lambda_{b}$&$-\sqrt{2}\lambda_{c}$&$0$&$0$&$0$\\
&$\bar{B}_{s}^{*}\Lambda_{c}$&$\frac{-\lambda_{b}}{2\sqrt{2}}$&$0$&$0$&$0$\\
\hline
\multirow{2}{*}{$C_{ij}(VB(3/2^{+}))$}&&$K^{*}\Xi_{bc}^{*}$&&&\\
&$K^{*}\Xi_{bc}^{*}$&$1$&&&\\
\bottomrule[1.00pt]
\bottomrule[1.00pt]
\end{tabular*}
\label{tab:bcqqs1}
\end{table}

By employing the aforementioned $S$-wave interaction as the kernel in the Bethe-Salpeter (BS) equation, we can derive the scattering amplitude for the coupled channels.
The on-shell factorized form of the BS equation is defined as \cite{Oset:1997it}
\begin{equation}
\begin{aligned}
T = [1-vG]^{-1}v,
\end{aligned}
\label{eq:BSE}
\end{equation}
where $v$ is the transition amplitude between coupled channels, defined by Eq. \eqref{eq:vij2} with the corresponding coefficients provided in Tables \ref{tab:ccqqs1}-\ref{tab:bcqqs1}, and $G$ represents the intermediate meson-baryon loop function.
The matrix $G$ is a diagonal matrix, and its diagonal elements can be expressed as

\begin{equation}
\begin{aligned}
G_{l}=i \int \frac{d^{4} q}{(2 \pi)^{4}}\frac{2M_{l}}{(P-q)^{2}-M_{l}^{2}+i \epsilon} \frac{1}{q^{2}-m_{l}^{2}+i \epsilon},
\end{aligned}
\label{eq:G}
\end{equation}
where the subscript $l$ denotes the label for the $l$-th channel.
Note that this $G_{l}$ function displays asymptotic logarithmic divergence.
There are two primary approaches to addressing this problem: namely, the three-momentum cutoff method \cite{Oset:1997it} and the dimensional regularization method \cite{Oller:2000fj,Jido:2003cb}.
Using the cutoff method, Eq. \eqref{eq:G} can be expressed in the following form
\begin{equation}
\begin{aligned}
G_{l}(s)=\int_{0}^{q_{max}} \frac{q^{2} d q}{2 \pi^{2}} \frac{1}{2 \omega_{l}(q)} \frac{M_{l}}{E_{l}(q)} \frac{1}{p^{0}+k^{0}-\omega_{l}(q)-E_{l}(q)+i \epsilon},
\end{aligned}
\label{eq:GCO}
\end{equation}
where $q=|\vec{q}|$, $p^{0} + k^{0} = \sqrt{s}$, $\omega_{l}(q) = \sqrt{q^{2} + m_{l}^{2}}$, and $E_{l}(q) = \sqrt{q^{2} + M_{l}^{2}}$, with $m_{l}$ and $M_{l}$ are the masses of the meson and baryon, respectively.
The $q_{\text{max}}$ is a theoretically free parameter, the three-momentum cutoff.
The expression for the dimensional regularization method is
\begin{equation}
\begin{aligned}
G_{l}(s)=& \frac{2 M_{l}}{16 \pi^{2}}\left\{a_{l}(\mu)+\ln \frac{M_{l}^{2}}{\mu^{2}}+\frac{m_{l}^{2}-M_{l}^{2}+s}{2 s} \ln \frac{m_{l}^{2}}{M_{l}^{2}}\right.\\
&+\frac{q_{cml}(s)}{\sqrt{s}}\left[\ln \left(s-\left(M_{l}^{2}-m_{l}^{2}\right)+2 q_{cml}(s) \sqrt{s}\right)\right.\\
&+\ln \left(s+\left(M_{l}^{2}-m_{l}^{2}\right)+2 q_{cml}(s) \sqrt{s}\right) \\
&-\ln \left(-s-\left(M_{l}^{2}-m_{l}^{2}\right)+2 q_{cml}(s) \sqrt{s}\right) \\
&\left.\left.-\ln \left(-s+\left(M_{l}^{2}-m_{l}^{2}\right)+2 q_{cml}(s) \sqrt{s}\right)\right]\right\},
\end{aligned}
\label{eq:GDR}
\end{equation}
which has one free parameter, the regularization scale $\mu$, and the subtraction constant $a_{l}(\mu)$ depends on the chosen $\mu$. 
The $q_{cml}(s)$ is the three-momentum of the particle in the center-of-mass frame:
\begin{equation}
\begin{aligned}
q_{cml}(s)=\frac{\lambda^{1 / 2}\left(s, M_{l}^{2}, m_{l}^{2}\right)}{2 \sqrt{s}},
\end{aligned}
\end{equation}
with the K\"all\'en triangle function $\lambda(a, b, c)=a^{2}+b^{2}+c^{2}-2(a b+a c+b c)$.

In this paper, we employ the dimensional regularization method to circumvent the singularity inherent in the three-momentum cutoff method above the threshold.
We use the empirical value of the regularization scale $\mu=800$ MeV as in Refs. \cite{Marse-Valera:2022khy,Roca:2024nsi}.
Then set $\mu=q_{\text{max}}$ and determine $a_{l}(\mu)$ for a given channel by matching the real part values of $G$ obtained through the above two renormalization methods at the threshold, following the approach described in Refs. \cite{Wang:2023mdj,Wang:2024yjp}
\begin{equation}
\begin{aligned}
a_{l}(\mu)=\frac{16\pi^{2}}{2M_{l}}\left[G^{CO}(s_{thr},q_{max})-G^{DR}(s_{thr},\mu)\right],
\end{aligned}
 \label{eq:ai}
\end{equation}
where $G^{CO}$ and $G^{DR}$ are given by Eqs. \eqref{eq:GCO} and \eqref{eq:GDR}, respectively.
This approach can minimize the number of theoretical free parameters and reduce uncertainty as much as possible in the absence of experimental data.

The two-body scattering amplitudes for the coupled channels can be calculated using Eq. \eqref{eq:BSE}.
We first look for peak structures in the diagonal amplitudes to investigate the dynamics of the generated resonances.
The masses and widths of the resonances are determined by the location of the poles of the scattering amplitudes on the complex Riemann sheets.
For a given channel, the loop function in the second Riemann sheet can be expressed in terms of the one in the first Riemann sheet by
\begin{equation}
\begin{aligned}
G_{l}^{(II)}(s)&=G_{l}^{I}(s)-2i \text{Im}G_{l}^{I}(s) \\
&=G_{l}^{I}(s)+\frac{i}{2\pi}\frac{M_{l}q_{cml}(s)}{\sqrt{s}},
\end{aligned}
\end{equation}
where $G_{l}^{I}(s)$ is given by Eq. \eqref{eq:GDR}.
When searching for poles, it is necessary to use $G_{l}^{I}(s)$ for $\text{Re}[\sqrt{s}] < m_{l} + M_{l}$ and switch to $G_{l}^{II}(s)$ for $\text{Re}[\sqrt{s}] > m_{l} + M_{l}$.

Furthermore, to evaluate the coupling of the generated state for a given channel, we can perform a Laurent expansion of the amplitude around the pole $\sqrt{s_{p}}$ on the complex plane \cite{Yamagata-Sekihara:2010kpd}
\begin{equation}
T_{ij}=\frac{g_{i}g_{j}}{\sqrt{s}-\sqrt{s_{p}}},
\end{equation}
where $g_{i}$ and $g_{j}$ are the couplings of the $i$-th and $j$-th channels, respectively.

\section{Numerical results}\label{sec:Results}

In Tables \ref{tab:ccqqs1}-\ref{tab:bcqqs1}, a positive coefficient corresponds to an attractive interaction in Eq. \eqref{eq:vij2}.
Given the magnitude of these coefficients, we expect the dynamic generation of resonances in the $K^{(*)}\Xi_{cc}^{(*)}$, $K^{(*)}\Xi_{bb}^{(*)}$, $K^{(*)}\Xi_{bc}^{(*)}$, and $K^{(*)}\Xi_{bc}^{'}$ channels.
We then present numerical results for the $ccqq\bar{s}$ system in \ref{sec:sector1}, the $bbqq\bar{s}$ system in \ref{sec:sector2}, and the $bcqq\bar{s}$ system in \ref{sec:sector3}.

\subsection{The $ccqq\bar{s}$ system}\label{sec:sector1}

\begin{figure}[htbp]
\begin{minipage}{0.49\linewidth}
\centering
\includegraphics[width=1\linewidth,trim=0 0 0 0,clip]{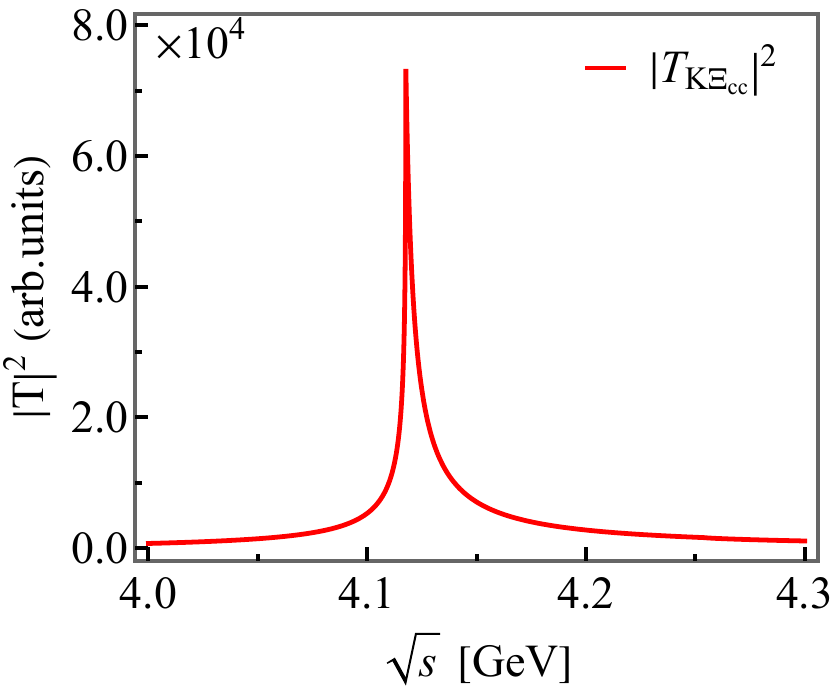}
\end{minipage}
\begin{minipage}{0.49\linewidth}
\centering
\includegraphics[width=1\linewidth,trim=0 0 0 0,clip]{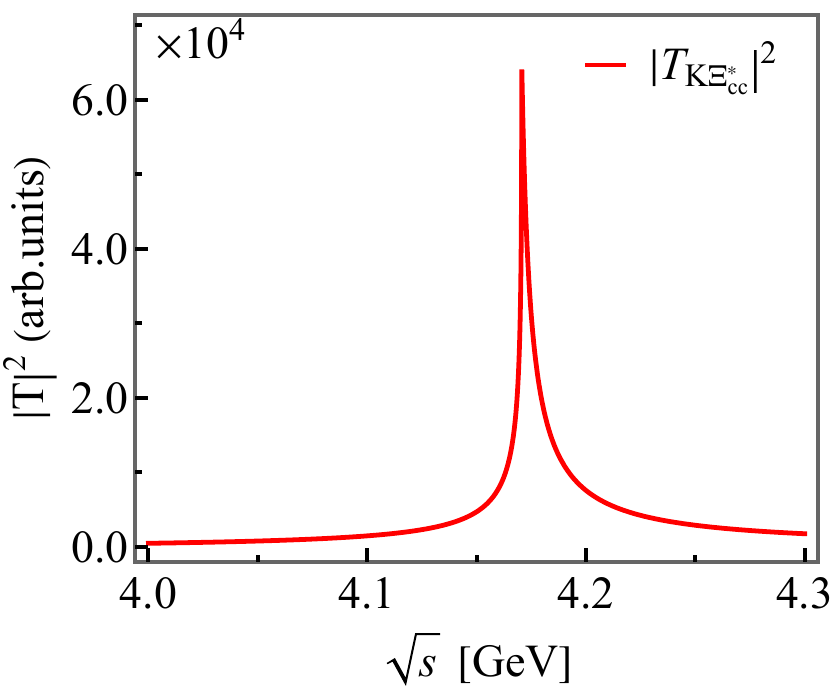}
\end{minipage}	
\begin{minipage}{0.49\linewidth}
\centering
\includegraphics[width=1\linewidth,trim=0 0 0 0,clip]{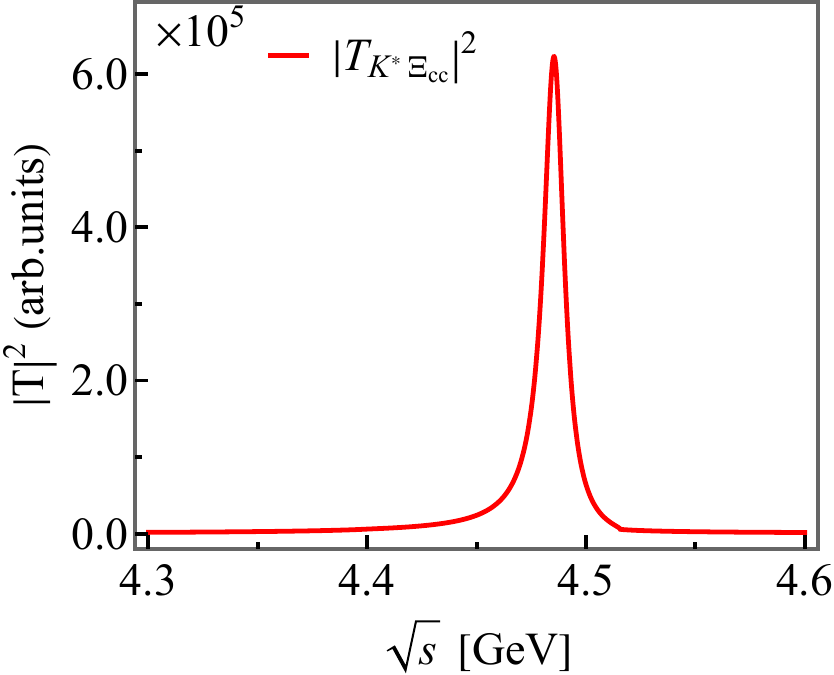}
\end{minipage}
\begin{minipage}{0.49\linewidth}
\centering
\includegraphics[width=1\linewidth,trim=0 0 0 0,clip]{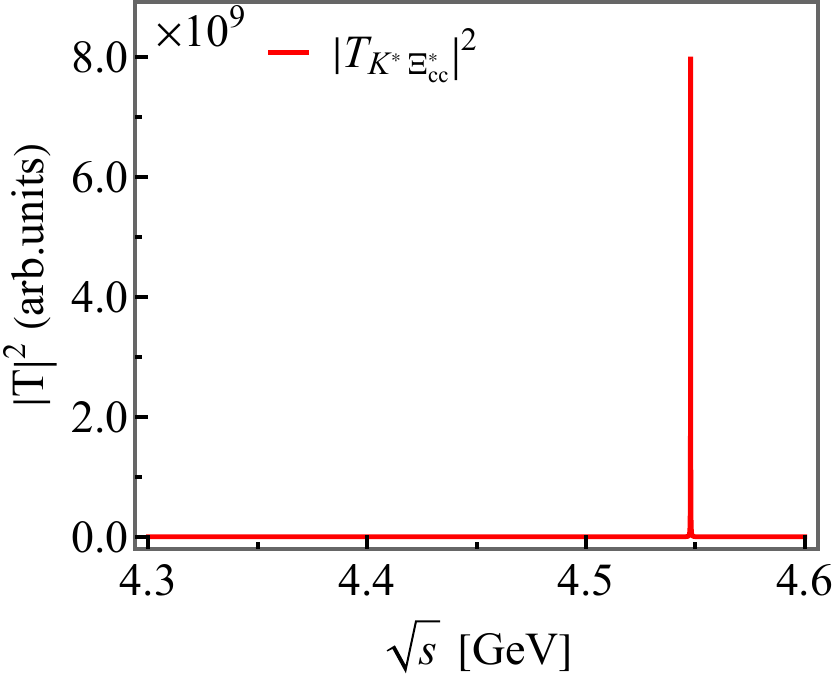}
\end{minipage}
\caption{The modulus square of the amplitudes in the $ccqq\bar{s}$ system.}
\label{fig:ccqqs}
\end{figure}

We first identify four peak structures in the $ccqq\bar{s}$ system, which are located within the amplitudes of the diagonal elements corresponding to $PB(1/2^{+})$, $PB(3/2^{+})$, $VB(1/2^{+})$, and $VB(3/2^{+})$, as shown in Fig. \ref{fig:ccqqs}.
Subsequently, we find four corresponding poles located on different complex energy planes, as listed in Table \ref{tab:ccqqs}.
The $4113.99$ MeV state, which lies $3.66$ MeV below the $K\Xi_{cc}$ channel threshold, primarily couples to this channel. This suggests that it could mostly qualify as a $K\Xi_{cc}$ molecular state with quantum numbers $I(J^{P}) = 0(1/2^{-})$.
It is a virtual state because the pole lies on the Riemann sheet $(-,+)$.
In the $PB(3/2^{+})$ sector, there is only one channel $K\Xi_{cc}^{*}$, and the state at $4166.59$ MeV is a virtual state dynamically generated in this single channel with a binding energy of $4.06$ MeV.

\begin{table}[htbp]
\centering
\renewcommand\tabcolsep{0.8mm}
\renewcommand{\arraystretch}{1.50}
\caption{The poles (in MeV) and their couplings for every channel in the $ccqq\bar{s}$ system. The symbols $+$ and $-$ correspond to the channel statuses in Table \ref{tab:coupledchannels}, where $+$ indicates that the corresponding channel is closed and $-$ indicates that it is open, following the same order as listed in Table \ref{tab:coupledchannels}. The largest coupling constant is denoted in bold.}
\begin{tabular*}{86mm}{@{\extracolsep{\fill}}lccc}
\toprule[1.00pt]
\toprule[1.00pt]
$I(J^{P})$&Poles position&\multicolumn{2}{c}{Couplings}\\
\hline
$0(\frac{1}{2}^{-})$&$4113.99$ $(-+)$&$|g_{K\Xi_{cc}}|=\bf{1.47}$&$|g_{D_{s}\Lambda_{c}}|=0.73$\\
$0(\frac{3}{2}^{-})$&$4166.59$ $(-)$&$|g_{K\Xi_{cc}^{*}}|=\bf{1.49}$&\\
$0(\frac{1}{2}^{-},\frac{3}{2}^{-})$&$4485.67-5.75i$ $(+-)$&$|g_{K^{*}\Xi_{cc}}|=\bf{2.14}$&$|g_{D_{s}^{*}\Lambda_{c}}|=0.56$\\
$0(\frac{1}{2}^{-},\frac{3}{2}^{-},\frac{5}{2}^{-})$&$4547.84$ $(+)$&$|g_{K^{*}\Xi_{cc}^{*}}|=\bf{1.94}$&\\
\bottomrule[1.00pt]
\bottomrule[1.00pt]
\end{tabular*}
\label{tab:ccqqs}
\end{table}

For the $VB(1/2^{+})$ case, we obtain a state at $4485.67-5.75i$ MeV that is degenerate in $J^{P}=1/2^{-}$ and $J^{P}=3/2^{-}$.
The couplings to the $K^{*}\Xi_{cc}$ and $D_{s}^{*}\Lambda_{c}$ channels are found to be $|g_{K^{*}\Xi_{cc}}|=2.14$ and $|g_{D_{s}^{*}\Lambda_{c}}|=0.56$, respectively. 
This suggests that the state is predominantly a quasi-bound of $K^{*}\Xi_{cc}$ channel, with a binding energy of approximately $30$ MeV.
It can decay into the $D_{s}^{*}\Lambda_{c}$ channel, contributing a width of $11.5$ MeV, which is twice the imaginary part of the pole.
For the $VB(3/2^{+})$ case, the $4547.84$ MeV state would be a $K^{*}\Xi_{cc}^{*}$ bound state with a binding energy about $20$ MeV, and degenerate in $J^{P}=1/2^{-}$, $J^{P}=3/2^{-}$ and $J^{P}=5/2^{-}$. 
Note that there is a significant difference between the peak structures of the first two virtual states and those of the last two bound or quasi-bound states in Fig. \ref{fig:ccqqs}.

For the $K\Xi_{cc}$ system, Ref. \cite{Sheng:2024hkf} demonstrated that considering the coupled channels interaction between $K\Xi_{cc}$ and $K^{*}\Xi_{cc}$ could form a loosely bound state when the cutoff $\Lambda$ is set to at least $1.45$ GeV within the one-boson-exchange model.
Although the interactions considered and the properties of the states differ from ours, the conclusion that molecular states can be obtained remains consistent with our findings.
Additionally, the aforementioned study identifies a bound state in the $K^{*}\Xi_{cc}$ system, with its binding energy determined by the theoretical free parameter $\Lambda$.
This finding is consistent with our results; however, we also reported a narrower width.

\subsection{The $bbqq\bar{s}$ system}\label{sec:sector2}

\begin{figure}[htbp]
\begin{minipage}{0.49\linewidth}
\centering
\includegraphics[width=1\linewidth,trim=0 0 0 0,clip]{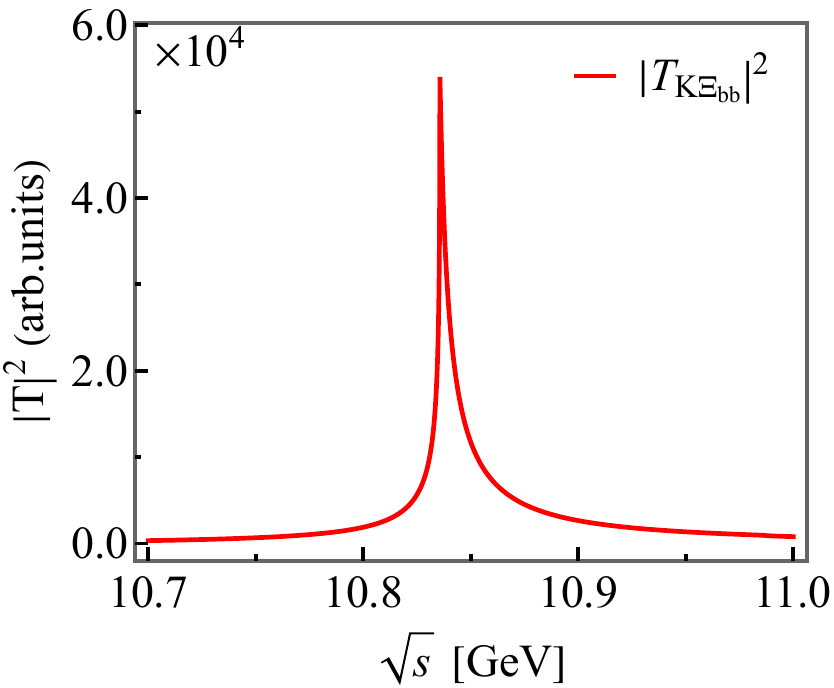}
\end{minipage}
\begin{minipage}{0.49\linewidth}
\centering
\includegraphics[width=1\linewidth,trim=0 0 0 0,clip]{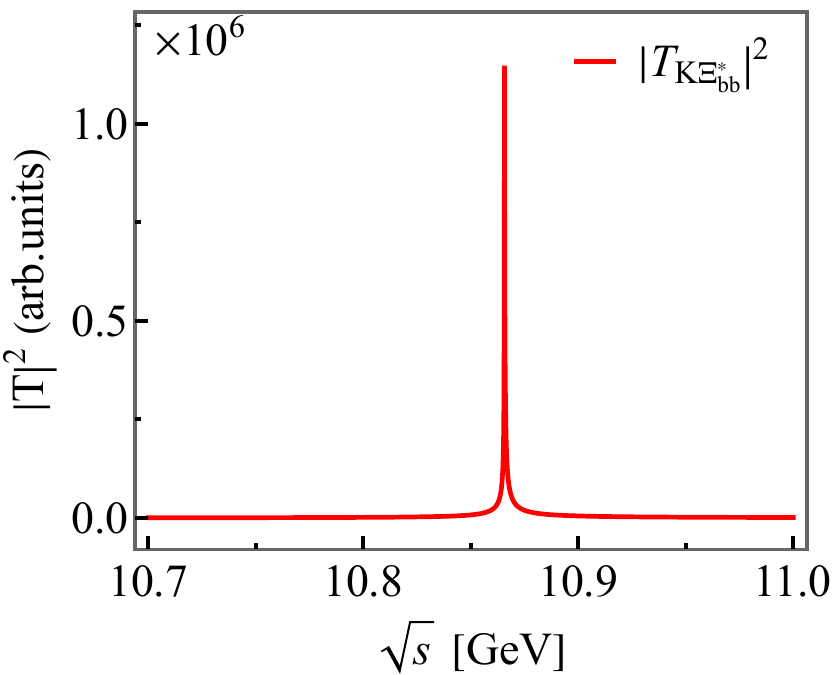}
\end{minipage}	
\begin{minipage}{0.49\linewidth}
\centering
\includegraphics[width=1\linewidth,trim=0 0 0 0,clip]{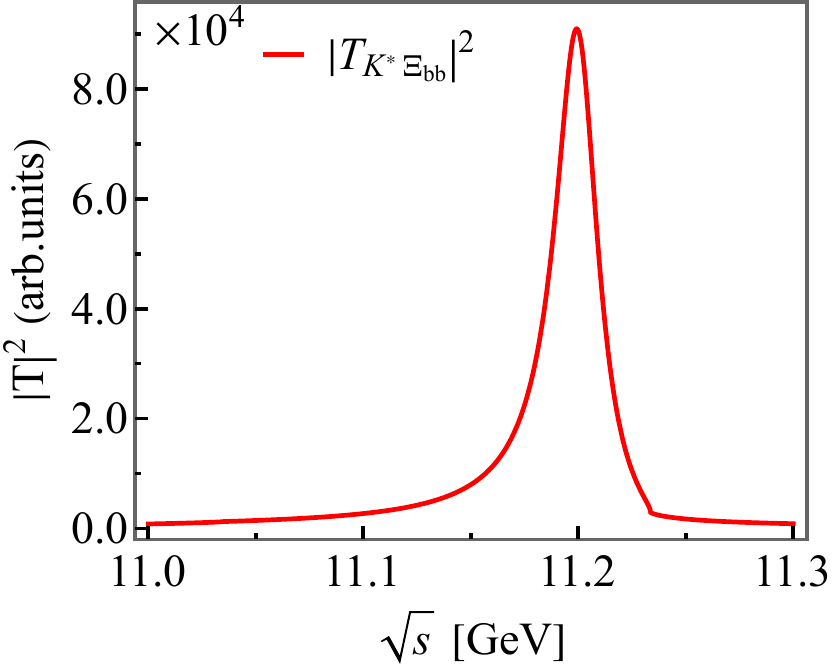}
\end{minipage}
\begin{minipage}{0.49\linewidth}
\centering
\includegraphics[width=1\linewidth,trim=0 0 0 0,clip]{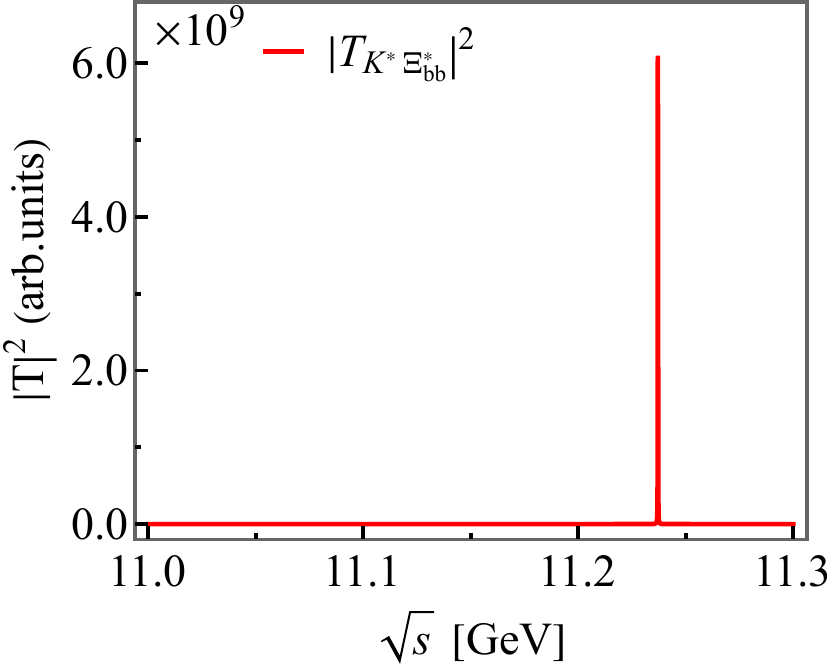}
\end{minipage}
\caption{The modulus square of the amplitudes in the $bbqq\bar{s}$ system.}
\label{fig:bbqqs}
\end{figure}

In the $bbqq\bar{s}$ system, we also obtain four candidates for molecular pentaquark states.
The peak structures and poles are presented in Fig. \ref{fig:bbqqs} and Table \ref{tab:bbqqs}, respectively.
The first two states at $10831.30$ MeV and $10865.53$ MeV are the virtual states formed by the $K\Xi_{bb}$ and $K\Xi_{bb}^{*}$ channels, respectively.
Similar to the $ccqq\bar{s}$ system, the binding energies of these two states are relatively small, at 4.34 MeV and 0.11 MeV.
The third state lies below the $K^{*}\Xi_{bb}$ threshold and corresponds to a quasi-bound state in this channel, with a binding energy of $33.31$ MeV.
This state has a width of approximately $\Gamma=24$ MeV, which originates from its decay into the $\bar{B}_{s}^{*}\Lambda_{b}$ channel.
The last one at $11237.08$ MeV with a zero width is a $K^{*}\Xi_{bb}^{*}$ bound state degenerate in $I(J^{P})=0(1/2^{-})$, $0(3/2^{-})$, and $0(5/2^{-})$.
Its binding energy is $26.53$ MeV.

\begin{table}[htbp]
\centering
\renewcommand\tabcolsep{0.8mm}
\renewcommand{\arraystretch}{1.50}
\caption{The poles (in MeV) and their couplings for every channel in the $bbqq\bar{s}$ system.}
\begin{tabular*}{86mm}{@{\extracolsep{\fill}}lccc}
\toprule[1.00pt]
\toprule[1.00pt]
$I(J^{P})$&Poles position&\multicolumn{2}{c}{Couplings}\\
\hline
$0(\frac{1}{2}^{-})$&$10831.30$ $(-+)$&$|g_{K\Xi_{bb}}|=\bf{1.53}$&$|g_{\bar{B}_{s}\Lambda_{b}}|=0.85$\\
$0(\frac{3}{2}^{-})$&$10865.53$ $(-)$&$|g_{K\Xi_{bb}^{*}}|=\bf{0.55}$&\\
$0(\frac{1}{2}^{-},\frac{3}{2}^{-})$&$11200.30-12.20i$ $(+-)$&$|g_{K^{*}\Xi_{bb}}|=\bf{1.94}$&$|g_{\bar{B}_{s}^{*}\Lambda_{b}}|=0.57$\\
$0(\frac{1}{2}^{-},\frac{3}{2}^{-},\frac{5}{2}^{-})$&$11237.08$ $(+)$&$|g_{K^{*}\Xi_{bb}^{*}}|=\bf{1.80}$&\\
\bottomrule[1.00pt]
\bottomrule[1.00pt]
\end{tabular*}
\label{tab:bbqqs}
\end{table}

It can be seen that the results obtained for the $bbqq\bar{s}$ system are similar to those obtained for the $ccqq\bar{s}$ system.
Among the four states obtained, the first two are virtual states with relatively small binding energies, while the last two are quasi-bound and bound states with binding energies of around $20-30$ MeV. 
These results reflect the symmetry of heavy quarks, since the two systems exhibit similar interactions as shown in Tables \ref{tab:ccqqs1} and \ref{tab:bbqqs1}.

\subsection{The $bcqq\bar{s}$ system}\label{sec:sector3}

\begin{figure}[htbp]
\begin{minipage}{0.49\linewidth}
\centering
\includegraphics[width=1\linewidth,trim=0 0 0 0,clip]{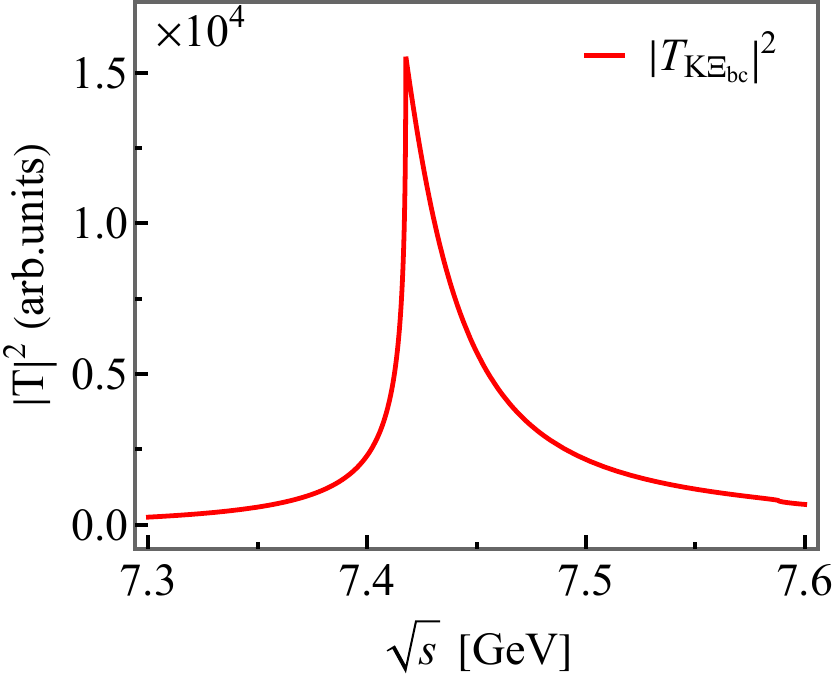}
\end{minipage}
\begin{minipage}{0.49\linewidth}
\centering
\includegraphics[width=1\linewidth,trim=0 0 0 0,clip]{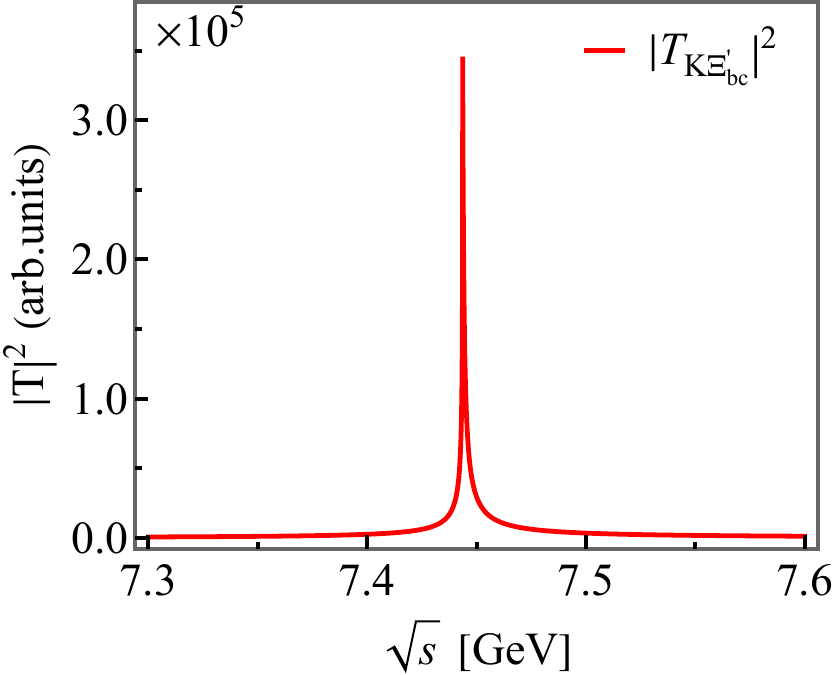}
\end{minipage}	
\begin{minipage}{0.49\linewidth}
\centering
\includegraphics[width=1\linewidth,trim=0 0 0 0,clip]{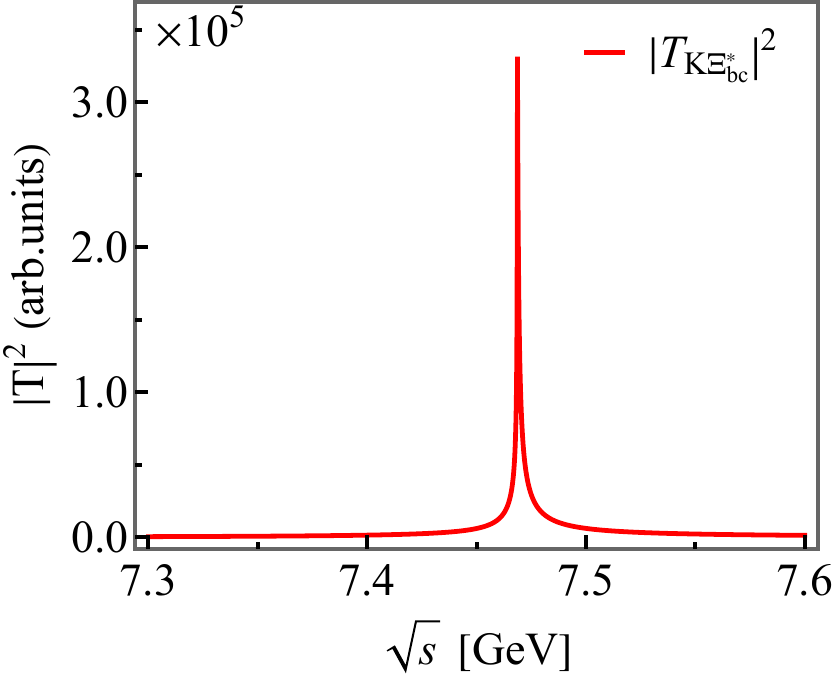}
\end{minipage}
\begin{minipage}{0.49\linewidth}
\centering
\includegraphics[width=1\linewidth,trim=0 0 0 0,clip]{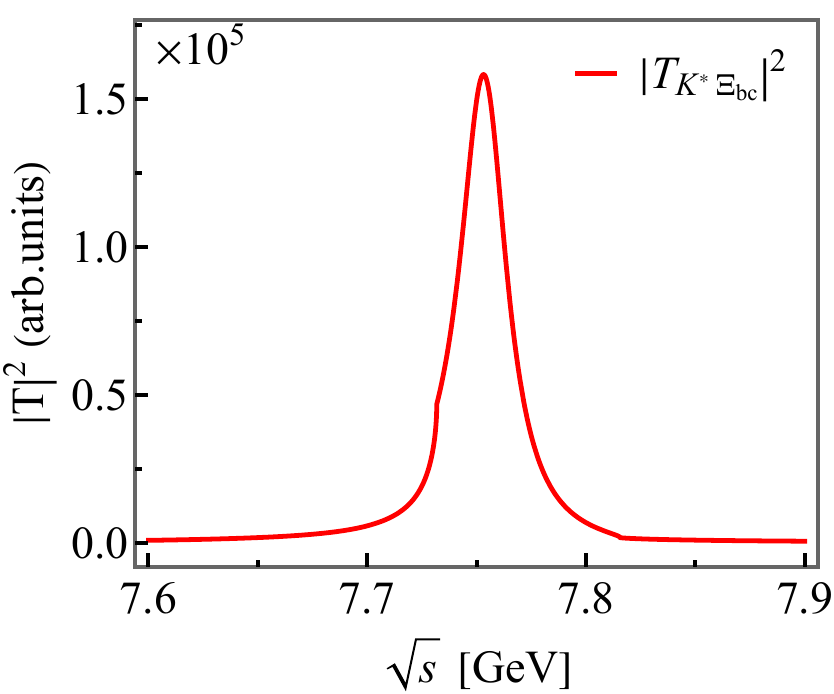}
\end{minipage}
\begin{minipage}{0.49\linewidth}
\centering
\includegraphics[width=1\linewidth,trim=0 0 0 0,clip]{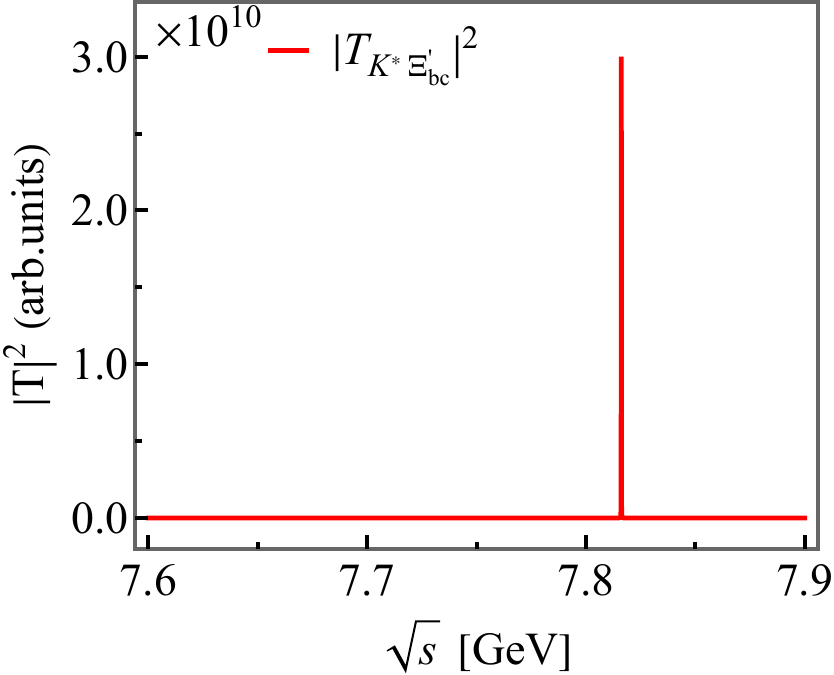}
\end{minipage}
\begin{minipage}{0.49\linewidth}
\centering
\includegraphics[width=1\linewidth,trim=0 0 0 0,clip]{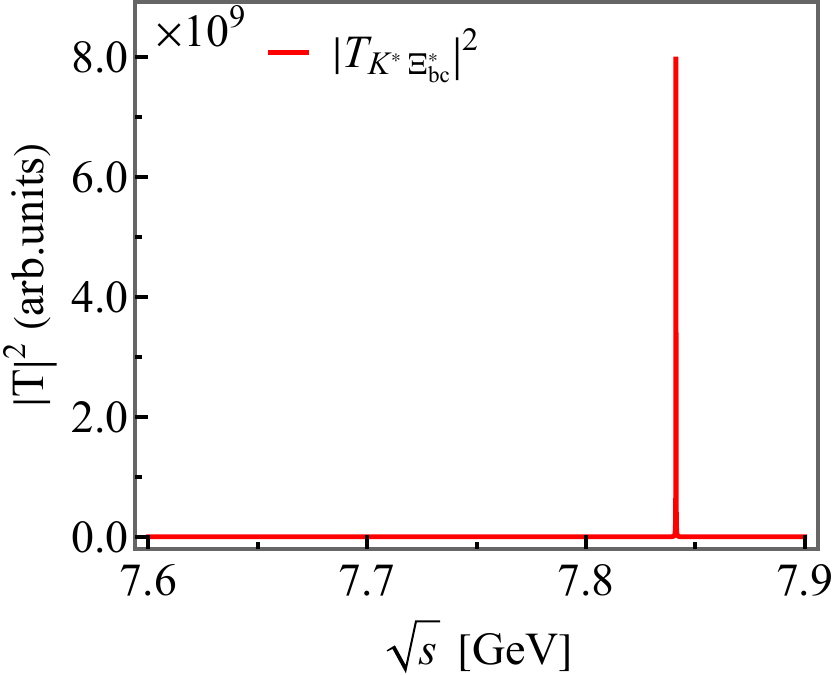}
\end{minipage}
\caption{The modulus square of the amplitudes in the $bcqq\bar{s}$ system.}
\label{fig:bcqqs}
\end{figure}

In this subsection, we conduct numerical analyses to investigate the $bcqq\bar{s}$ system.
Figure \ref{fig:bcqqs} displays the modulus square of the amplitudes, whereas Table \ref{tab:bcqqs} presents the identified poles.
For the case of $PB(1/2^{+})$, we find two states with $I(J^{P})=0(1/2^{-})$ located below the $K\Xi_{bc}$ and $K\Xi_{bc}^{'}$ thresholds, respectively.
Although the pole of the first state resides on the second Riemann sheet, its mass lies $22.44$ MeV below the $K\Xi_{bc}$ channel threshold.
It is puzzling that the state exhibits a relatively large width of $\Gamma=50$ MeV, which is twice the imaginary part of the pole, given that it lies below the thresholds of the four channels, $K\Xi_{bc}$, $K\Xi_{bc}^{'}$, $D_{s}\Lambda_{b}$, and $\bar{B}_{s}\Lambda_{c}$.
Additionally, it is about $200$ MeV below the $D_{s}\Lambda_{b}$ channel but has a stronger coupling to the $D_{s}\Lambda_{b}$ channel compared to the $K\Xi_{bc}$ channel, as evidenced by coupling strengths of $|g_{D_{s}\Lambda_{b}}|=3.03$ versus $|g_{K\Xi_{bc}}|=2.91$.
The second state at $7443.07$ MeV couples exclusively to the $K\Xi_{bc}^{'}$ channel and undoubtedly represents a virtual state formed by this channel, with a binding energy of $0.58$ MeV.

\begin{table*}[htbp]
\centering
\renewcommand\tabcolsep{2mm}
\renewcommand{\arraystretch}{1.50}
\caption{The poles (in MeV) and their couplings for every channel in the $bcqq\bar{s}$ system.}
\begin{tabular*}{178mm}{@{\extracolsep{\fill}}lccccc}
\toprule[1.00pt]
\toprule[1.00pt]
$I(J^{P})$&Poles position&\multicolumn{4}{c}{Couplings}\\
\hline
\multirow{2}{*}{$0(\frac{1}{2}^{-})$}&$7395.21-26.26i$ $(-+++)$&$|g_{K\Xi_{bc}}|=2.91$&$|g_{K\Xi_{bc}^{'}}|=0.00$&$|g_{D_{s}\Lambda_{b}}|=\bf{3.03}$&$|g_{\bar{B}_{s}\Lambda_{c}}|=0.74$\\
&$7443.07$ $(+-++)$&$|g_{K\Xi_{bc}}|=0.00$&$|g_{K\Xi_{bc}^{'}}|=\bf{0.84}$&$|g_{D_{s}\Lambda_{b}}|=0.00$&$|g_{\bar{B}_{s}\Lambda_{c}}|=0.00$\\
\hline
$0(\frac{3}{2}^{-})$&$7468.07$ $(-)$&$|g_{K\Xi_{bc}^{*}}|=\bf{0.84}$&&&\\
\hline
\multirow{2}{*}{$0(\frac{1}{2}^{-},\frac{3}{2}^{-})$}&$7753.79-13.02i$ $(++--)$&$|g_{K^{*}\Xi_{bc}}|=\bf{2.34}$&$|g_{K^{*}\Xi_{bc}^{'}}|=0.00$&$|g_{D_{s}^{*}\Lambda_{b}}|=0.89$&$|g_{\bar{B}_{s}^{*}\Lambda_{c}}|=0.19$\\
&$7816.16$ $(-+--)$&$|g_{K^{*}\Xi_{bc}}|=0.00$&$|g_{K^{*}\Xi_{bc}^{'}}|=\bf{1.86}$&$|g_{D_{s}^{*}\Lambda_{b}}|=0.00$&$|g_{\bar{B}_{s}^{*}\Lambda_{c}}|=0.00$\\
\hline
$0(\frac{1}{2}^{-},\frac{3}{2}^{-},\frac{5}{2}^{-})$&$7841.15$ $(+)$&$|g_{K^{*}\Xi_{bc}^{*}}|=\bf{1.86}$&&&\\
\bottomrule[1.00pt]
\bottomrule[1.00pt]
\end{tabular*}
\label{tab:bcqqs}
\end{table*}

\begin{figure}[htbp]
\begin{minipage}{0.49\linewidth}
\centering
\includegraphics[width=1\linewidth,trim=0 0 0 0,clip]{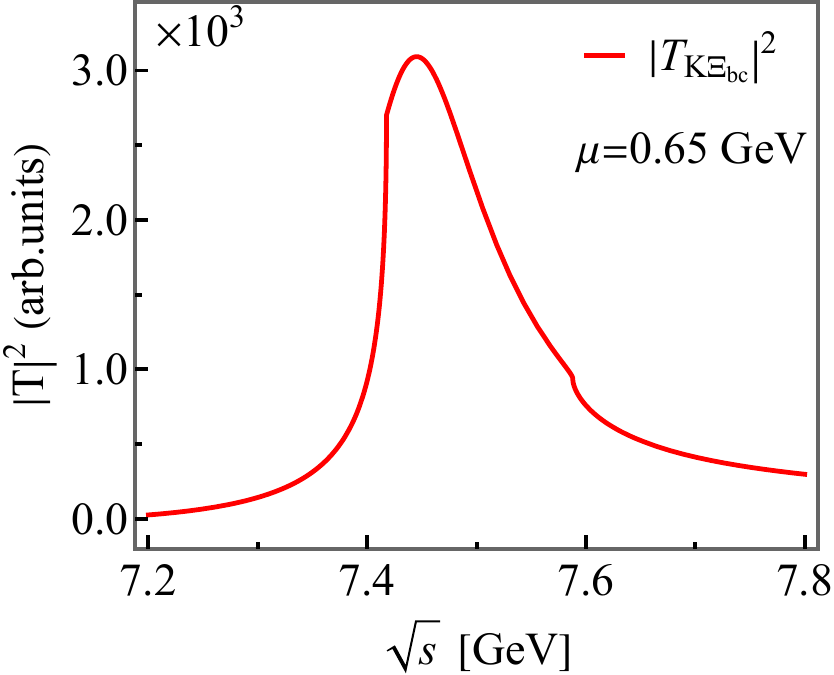}
\end{minipage}
\begin{minipage}{0.49\linewidth}
\centering
\includegraphics[width=1\linewidth,trim=0 0 0 0,clip]{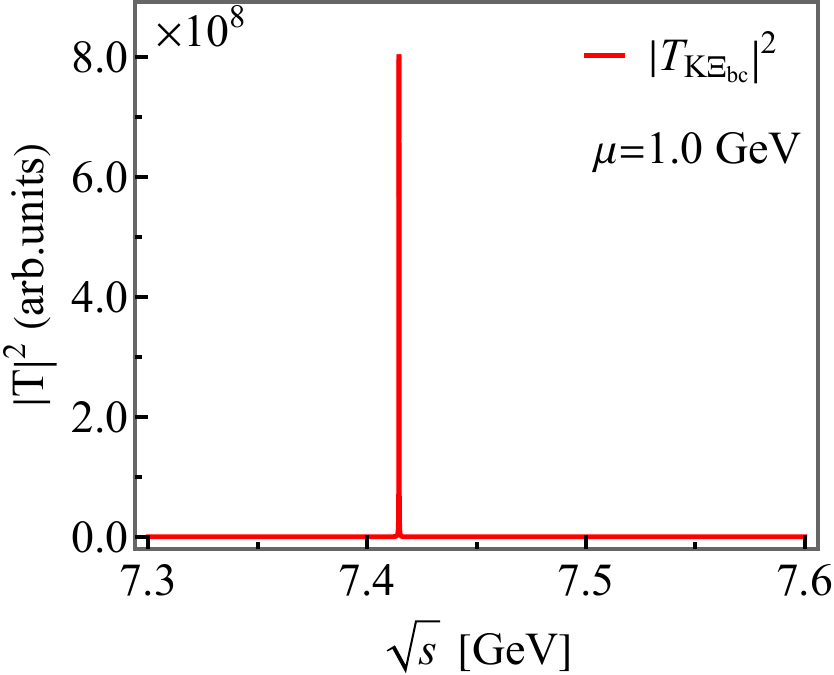}
\end{minipage}	
\caption{The modulus square of the amplitudes in the $K\Xi_{bc}$ coupled channels system are shown for $\mu=0.65$ GeV (left panel) and $\mu=1.0$ GeV (right panel), respectively.}
\label{fig:bcqqs6501000}
\end{figure}

To gain deeper insights into the first state at $7395.21-26.26i$ MeV within the $K\Xi_{bc}$ coupled channels system, we examine how its properties vary with changes in the regularization scale $\mu$.
If the free parameter $\mu$ is set to 650 MeV, the pole of this state is located at $7432.44-75.63i$ MeV, corresponding to a resonance lying above the $K\Xi_{bc}$ threshold $7417.64$ MeV, which enables its decay into the $K\Xi_{bc}$ channel.
But the coupling to $D_{s}\Lambda_{b}$ remains stronger than that to $K\Xi_{bc}$.
When we set $\mu=1000$ MeV, the pole moves to $7414.54$ MeV on the Riemann sheet $(++++)$.
This pole has a mass lower than the $K\Xi_{bc}$ threshold, with $|g_{K\Xi_{bc}}|=1.13$ being greater than $|g_{D_{s}\Lambda_{b}}|=0.94$.
The corresponding modulus square of the amplitudes are presented in Fig. \ref{fig:bcqqs6501000}.
Therefore, we propose that this state is a bound state of the $K\Xi_{bc}$ channel with a binding energy of $3.10$ MeV.
Of course, it is critically dependent on the free parameter $\mu$ within the theoretical framework.

In the case of $PB(3/2^{+})$, the pole at $7468.07$ MeV would be mostly a $K\Xi_{bc}^{*}$ virtual state with a binding energy of $0.57$ MeV.
In the case of $VB(1/2^{+})$, the two resulting states are degenerate with $J^{P}=1/2^{-}$ and $J^{P}=3/2^{-}$.
The one of $7753.79-13.02i$ MeV is a molecule of $K^{*}\Xi_{bc}$ as it exhibits the strongest coupling to this channel and can decay into the open channels $D_{s}^{*}\Lambda_{b}$ and $\bar{B}_{s}^{*}\Lambda_{c}$.
The other one at $7816.16$ MeV would be a molecular $K^{*}\Xi_{bc}^{'}$ state bound by about $25.45$ MeV with zero width, which does not couple to other channels in this sector.
For the case $VB(3/2)^{+}$, we find one bound state that is degenerate with $J^{P} = 1/2^{-}$, $3/2^{-}$, and $5/2^{-}$ below the $K^{*}\Xi_{bc}^{*}$ threshold.
It locates at around $7841.15$ MeV, so its binding energy is about $25.46$ MeV with a zero width. 

\begin{table*}[htbp]
\centering
\renewcommand\tabcolsep{2mm}
\renewcommand{\arraystretch}{1.50}
\caption{The positions of the poles vary with the regularization scale $\mu$ (in MeV).}
\begin{tabular*}{178mm}{@{\extracolsep{\fill}}lccccc}
\toprule[1.00pt]
\toprule[1.00pt]
$I(J^{P})$&$\mu=650$&$\mu=750$&$\mu=850$&$\mu=950$&$\mu=1050$\\
\hline
&\multicolumn{5}{c}{$ccqq\bar{s}$ system} \\
\hline
$0(\frac{1}{2}^{-})$&$4076.95$&$4108.32$&$4116.81$&$4116.98$&$4112.38$\\
\hline
$0(\frac{3}{2}^{-})$&$4144.60$&$4162.09$&$4169.25$&$4170.54$&$4167.93$\\
\hline
$0(\frac{1}{2}^{-},\frac{3}{2}^{-})$&$4505.36-4.10i$&$4493.21-5.33i$&$4477.24-6.03i$&$4457.97-6.14i$&$4435.91-5.50i$\\
\hline
$0(\frac{1}{2}^{-},\frac{3}{2}^{-},\frac{5}{2}^{-})$&$4562.76$&$4553.73$&$4541.12$&$4525.42$&$4507.11$\\
\hline
&\multicolumn{5}{c}{$bbqq\bar{s}$ system} \\
\hline
$0(\frac{1}{2}^{-})$&$10797.57-42.51i$&$10823.27$&$10834.72$&$10834.91$&$10830.09$\\
\hline
$0(\frac{3}{2}^{-})$&$10856.04$&$10864.29$&$10865.46$&$10862.40$&$10856.49$\\
\hline
$0(\frac{1}{2}^{-},\frac{3}{2}^{-})$&$11220.30-8.65i$&$11207.75-11.14i$&$11192.13-13.14i$&$11173.82-14.64i$&$11153.15-15.57i$\\
\hline
$0(\frac{1}{2}^{-},\frac{3}{2}^{-},\frac{5}{2}^{-})$&$11253.33$&$11243.20$&$11230.31$&$11215.03$&$11197.72$\\
\hline
&\multicolumn{5}{c}{$bcqq\bar{s}$ system} \\
\hline
\multirow{2}{*}{$0(\frac{1}{2}^{-})$}&$7432.44-75.63i$&$7407.45-47.55i$&$7410.84$&$7417.22$&$7410.18$\\
&$7430.75$&$7441.06$&$7443.64$&$7441.57$&$7436.36$\\
\hline
$0(\frac{3}{2}^{-})$&$7455.79$&$7466.07$&$7468.64$&$7466.56$&$7461.34$\\
\hline
\multirow{2}{*}{$0(\frac{1}{2}^{-},\frac{3}{2}^{-})$}&$7784.75-15.35i$&$7764.98-14.64i$&$7741.78-9.91i$&$7727.14-0.19i$&$7708.63-0.16i$\\
&$7832.45$&$7822.38$&$7809.24$&$7793.45$&$7775.39$\\
\hline
$0(\frac{1}{2}^{-},\frac{3}{2}^{-},\frac{5}{2}^{-})$&$7857.43$&$7847.36$&$7834.22$&$7818.44$&$7800.38$\\
\bottomrule[1.00pt]
\bottomrule[1.00pt]
\end{tabular*}
\label{tab:poles}
\end{table*}

\begin{figure}[htbp]
\centering
\includegraphics[width=0.8\linewidth,trim=40 10 50 0,clip]{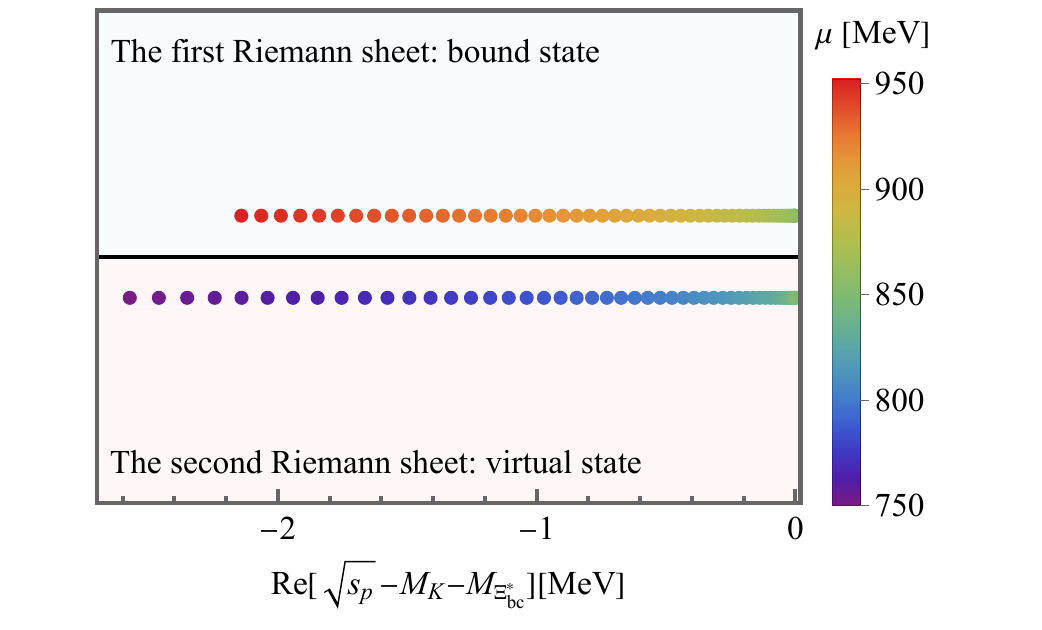}
\caption{Pole position of $\text{Re}[\sqrt{s_{p}} - M_{K} - M_{\Xi_{bc}^{*}}]$ for the $K\Xi_{bc}^{*}$ hadronic molecule as a function of the regularization scale $\mu$ from 750 to 950 MeV.}
\label{fig:F6}
\end{figure}

Finally, we present the pole positions of the hadronic molecular states with respect to the regularization scale $\mu$ ranging from $650$ MeV to $1050$ MeV in Table \ref{tab:poles}.
As the parameter $\mu$ increases, the virtual state undergoes a gradual transformation into a bound state.
For bound or quasi-bound states, the binding energy increases with increasing $\mu$.
As an example, Fig. \ref{fig:F6} presents a detailed analysis of the $\mu$-dependence of the pole position for the $K\Xi_{bc}^{*}$ system. 
This indicates that near $\mu=850$ MeV, the pole approaches the $K\Xi_{bc}^{*}$ threshold, which triggers significant changes in its dynamical properties and a transition between bound and virtual states. 
In general, the existence of a bound or virtual state and the binding energy are determined by the magnitude of the interactions.
Indeed, for the $T$ matrix in Eq. \eqref{eq:BSE} to have a pole, we require $\text{det}(1-vG)=0$. 
Therefore, the pole position cannot be determined solely by the interaction potential $v$ in our model, since it also depends on the loop function $G$.
The value of the regularization parameter cannot be theoretically established.
In conclusion, these hadronic molecular states can be dynamically generated within the empirically determined range of the parameter $\mu$, with values such as $\mu=650$ MeV and $\mu=1000$ MeV being widely employed in charm and bottom sectors \cite{Debastiani:2017ewu,Dias:2018qhp,Yu:2018yxl,Yu:2019yfr,Wang:2022aga,Dong:2021bvy,Lin:2023iww}.

\section{Summary}\label{sec:Summary}

Since the discovery of candidate hidden-charm pentaquark states $P_{c}$ and $P_{cs}$ in experiments, the search for new hadrons has spurred ongoing theoretical and experimental efforts.
Notably, the LHCb Collaboration has announced plans to investigate new hadronic states through upgraded detection methodologies.
This inspires us to theoretically study the double-heavy pentaquark states, which are currently under investigation by the LHCb Collaboration.

In this paper, we systematically investigate potential molecular pentaquark states with the quark configurations $ccqq\bar{s}$, $bbqq\bar{s}$, and $bcqq\bar{s}$.
The extended local hidden gauge approach is employed to derive the $S$-wave transition potentials from the meson-baryon interactions.
We then identify the dynamically generated molecular pentaquark states by extracting the poles of the amplitudes on the complex energy plane, which are obtained by solving the Bethe-Salpeter equation in coupled channels.
For the $ccqq\bar{s}$ system, we obtain four molecular states at the parameter $\mu=800$ MeV, including two virtual states, one quasi-bound state, and one bound state.
They are mainly coupled to the $K\Xi_{cc}$, $K\Xi_{cc}^{*}$, $K^{*}\Xi_{cc}$, and $K^{*}\Xi_{cc}^{*}$ channels, respectively.
For the $bbqq\bar{s}$ system, we obtain four molecular states, which are mainly coupled to the $K\Xi_{bb}$, $K\Xi_{bb}^{*}$, $K^{*}\Xi_{bb}$, and $K^{*}\Xi_{bb}^{*}$ channels, respectively.
They are all narrow states with binding energies in the order of $0.1$ to $33$ MeV.
For the $bcqq\bar{s}$ system, we obtain six molecular states, which strongly couple to the $K\Xi_{bc}$, $K\Xi_{bc}^{'}$, $K\Xi_{bc}^{*}$, $K^{*}\Xi_{bc}$, $K^{*}\Xi_{bc}^{'}$, and $K^{*}\Xi_{bc}^{*}$ channels, respectively.
Their quantum numbers are assigned as follows: $I(J^{P})=0(1/2^{-})$ for the $PB(1/2^{+})$ block, $I(J^{P})=0(3/2^{-})$ for the $PB(3/2^{+})$ block, while degeneracy occurs in $I(J^{P})=0(1/2^{-})$ and $0(3/2^{-})$ for the $VB(1/2^{+})$ block, and further degeneracy appears in $I(J^{P})=0(1/2^{-})$, $0(3/2^{-})$, and $0(5/2^{-})$ for the $VB(3/2^{+})$ block.
It is important to emphasize that our findings critically depend on the theoretically free parameter $\mu$, yet they remain valid across empirically plausible values of $\mu$.
Our results contribute to the spectroscopy of molecular pentaquark states.

\section*{Acknowledgements}

We would like to thank Professor Eulogio Oset for valuable comments.
This work is supported by the Natural Science Special Research Foundation of Guizhou University under Grant No. 2024028, and the National Natural Science Foundation of China under Grant No. 12265007.

 \addcontentsline{toc}{section}{References}

\end{document}